\documentclass[
reprint,
nofootinbib,
 amsmath,amssymb,
aps,
]{revtex4-1}
\usepackage{graphicx}% Include figure files
\usepackage{dcolumn}% Align table columns on decimal point
\usepackage{bm}% bold math
\usepackage{amssymb}
\newcommand{\ba}{\begin{eqnarray}}
\newcommand{\ea}{\end{eqnarray}}

\newcommand{\B}{{\cal{B}}}

\newcommand{\W}{{\cal {W}}}
\newcommand{\bbq}{\begin{quote}}
\newcommand{\eeq}{\end{quote}}

\newcommand{\tbb}{t_{\textrm{\tiny{bb}}}}

\newcommand{\tcoll}{t_{\textrm{\tiny{coll}}}}

\newcommand{\RR}{{}^{(3)}{\cal{R}}}

\newcommand{\CC}{{\cal{C}}}
\newcommand{\EE}{{\cal{E}}}
\newcommand{\FF}{{\cal{F}}}

\newcommand{\JJ}{{\cal{J}}}

\newcommand{\HH}{{\cal{H}}}
\newcommand{\KK}{{\cal{K}}}

\newcommand{\Y}{{\cal{Y}}}

\newcommand{\da}{\delta^{(A)}}
\newcommand{\dH}{\delta^{(\HH)}}
\newcommand{\dKK}{\delta^{(\KK)}}
\newcommand{\drho}{\delta^{(\rho)}}

\newcommand{\dd}{{\hbox{d}}}

\newcommand{\bW}{{\rm{\bf W}}}
\newcommand{\rtv}{r_{\rm{tv}}}

\newcommand{\Pspm}{\Psi_{2\pm}}
\newcommand{\Psp}{\Psi_{2+}}
\newcommand{\Psm}{\Psi_{2-}}
\newcommand{\Altb}{A_{(\textrm{\tiny{lt}})}}

\newcommand{\Sigltb}{\Sigma_{(\textrm{\tiny{lt}})}}
\newcommand{\Pltb}{\Psi_2^{(\textrm{\tiny{lt}})}}

\newcommand{\Ommi}{\Omega_{q0} ^{(m)}}

\newcommand{\OmLi}{\Omega_{q0} ^{(\Lambda)}}

\newcommand{\Omki}{\Omega_{q0} ^{(k)}}
\newcommand{\DDa}{{\textrm{\bf{D}}}^{(A)}}
\newcommand{\DDaltb}{{\textrm{\bf{D}}}^{(A)}_{(\textrm{\tiny{lt}})}}
\newcommand{\DDrho}{{\textrm{\bf{D}}}^{(\rho)}}
\newcommand{\DDrholtb}{{\textrm{\bf{D}}}^{(\rho)}_{(\textrm{\tiny{lt}})}}
\newcommand{\DDKK}{{\textrm{\bf{D}}}^{(\KK)}}
\newcommand{\DDKKltb}{{\textrm{\bf{D}}}^{(\KK)}_{(\textrm{\tiny{lt}})}}
\newcommand{\DDh}{{\textrm{\bf{D}}}^{(\HH)}}
\newcommand{\DDhltb}{{\textrm{\bf{D}}}^{(\HH)}_{(\textrm{\tiny{lt}})}}

\begin{document}

%\preprint{}

\title{Multiple nonspherical structures from the extrema of Szekeres scalars.}


\author{Roberto A. Sussman}
\email[]{sussman@nucleares.unam.mx}
%\thanks{}
\affiliation{Instituto de Ciencias Nucleares, Universidad Nacional Aut\'onoma de M\'exico (ICN-UNAM),
A. P. 70--543, 04510 M\'exico D. F., M\'exico.}

\author{I. Delgado Gaspar}
\email[]{ismael.delgadog@uaem.edu.mx}
%\thanks{}
\affiliation{Instituto de Investigaci\'on en Ciencias B\'asicas y Aplicadas, Universidad Aut\'onoma del Estado de Morelos, Av Universidad 1002, 62210, Cuernavaca, Morelos, M\'exico.}

\date{\today}

\begin{abstract}
We examine the spatial extrema (local maxima, minima and saddle points) of the covariant scalars (density, Hubble expansion, spatial curvature and eigenvalues of the shear and electric Weyl tensors) of the quasi--spherical Szekeres dust models. Sufficient conditions are obtained for the existence of distributions of multiple extrema in spatial comoving locations that can be prescribed through initial conditions. These distributions evolve without shell crossing singularities at least for ever expanding models (with or without cosmological constant) in the full evolution range where the models are valid. By considering the local maxima and minima of the density, our results allow for setting up elaborated networks of ``pancake'' shaped evolving cold dark matter over--densities and density voids whose spatial distribution and amplitudes can be controlled from initial data compatible with standard early Universe initial conditions. We believe that these results have an enormous range of potential application by providing a fully relativistic non--perturbative coarse grained modelling of cosmic structure at all scales.    
\end{abstract}
\pacs{04.20.Jb, 98.80.Jk, 98.65.Dx, 04.40.-b}
%\keywords{}

\maketitle


\section{Introduction}\label{intro}

The current era of ``Precision Cosmology'' has produced a large amount of high quality observational data at all astrophysical and cosmic scales whose theoretical interpretation requires a robust modelling of self--gravitating systems. Conventionally, the large cosmic scale dynamics of these sources is examined through linear perturbations on a $\Lambda$CDM background \cite{Liddle,malikwand,EMM}, while Newtonian gravity (perturbative and non--perturbative \cite{Padma}, as well as numerical simulations \cite{simu1,simu2}) is often employed for self--gravitational systems at smaller galactic and galactic cluster scales . 

A non--perturbative approach by means of analytic or numerical solutions of Einstein's equations \cite{EMM,kras1,kras2,BKHC2009} is less favoured to analyse observations because (unless powerful numerical methods are employed) the high non--linear complexity of Einstein's equations renders realistic models mathematically untractable. As a consequence, extremely idealised toy models are used in most cosmological applications  based on fully relativistic and non--perturbative methods. The most prominent example are the spherically symmetric Lema\^\i tre--Tolman (LT) models \cite{Lema1933,Tolm1934,Bond1947} (see extensive reviews in \cite{EMM,kras1,kras2,BKHC2009}), which have been used to describe fully relativistic non--perturbative evolution of cold dark matter (CDM) sources at all scales: from local collapsing structures of galactic scale embedded in an FLRW background (spherical collapse and ``top hat'' models \cite{Padma,BKHC2009}) to Gpc sized cosmic voids in the recent effort to explore the possibility of fitting large scale cosmological observations without the assumption of a dark energy source or a cosmological constant \cite{BCK2011,marranot,bisnotval}. 

Even if admitting that LT void models have failed to fit large scale observations \cite{redlich,zibin1,zibin2,bull,finns}, non--perturbative general relativistic models are still needed and can be useful to probe structure formation scenarios and to provide theoretical support to cosmological observations within the current $\Lambda$CDM paradigm. However, less idealised models that are not restricted by spherical symmetry are required for this purpose, as the CDM structures we observe in all scales (from galactic surveys \cite{cosmography1,cosmography2}) is far from spherically symmetric. In this context, an improvement on the limitations of LT models is furnished by the well known Szekeres solutions \cite{S75,Sz75,GoWa1982} (see their derivation and classification in  \cite{kras1,kras2,BKHC2009,BCK2011}), which are still restrictive but are endowed with more degrees of freedom. While an extensive literature already exists on the usage of Szekeres models to address various problems in cosmic structure modelling and observations fitting  \cite{
Bsz1,Bsz2,IRGW2008,Bole2009-cmb,KrBo2010,Bole2010-sn,BoCe2010,NIT2011,MPT,
PMT,BoSu2011,sussbol,WH2012,Buckley,Vrba,kokhan1,kokhan2}, their full potential for theoretical and empirical applications to Cosmology is still open for future development.         

A specific issue in which Szekeres solutions (specially quasi--spherical models of class I \cite{kras1,kras2,BKHC2009}) are particularly helpful is the modelling of cosmic structure. Previous work on this issue \cite{Bsz1,Bsz2,PMT,BoSu2011,sussbol,WH2012,Buckley,kokhan1,kokhan2} (see review in \cite{BKHC2009}) has addressed the study of various aspects of non--spherical sources, allowing for a coarse grained description of CDM structures currently observed (see specially \cite{Bsz1,Bsz2,BoSu2011}), which typically consist of spatial distributions of spheroidal under--dense regions (voids) of typical 30--50 Mpc size surrounded by a web of elongated filamentary over--dense regions \cite{struct}. 

Since a distribution of over--densities and voids can be identified with a distribution of spatial maxima and minima of the density, we aim in this work (a complementary related work is found in \cite{nuevo}) at extending and improving this literature by undertaking a comprehensive study of the spatial extrema (local maxima, minima and saddle points) of all Szekeres covariant scalars (not just the density). 

By working in spatial spherical coordinates, we show how the conditions for the existence and location of these extrema can be split into interdependent angular and radial conditions, the latter requiring  to be handled numerically. However, by considering Szekeres models compatible with a time preserved Periodic Local Homogeneity property based on the existence of a sequence of local homogeneity 2--spheres (analogous to  localised FLRW backgrounds), we are able to find rigorous sufficient conditions for the existence of distributions of arbitrary numbers of these spatial extrema. 

We also examine how spatial extrema can arise by means of a ``simulated shell crossing'' mechanism for generic models that do not admit this property. We discuss extensively the classification of the extrema, the conditions for their avoidance of shell crossings and present a procedure to specify their spatial location at all times from selected initial conditions. 

These results allow us to set up elaborated networks of an arbitrary number of evolving pancake--like over--densities and density voids, in which the spatial density maxima or minima can be placed in prescribed comoving spatial locations at all times by realistic initial data. We believe that such evolving networks can provide a non--trivial coarse grained description of observed structures in a wide range of astrophysical and cosmological scales (specially in the supercluster scale \cite{superstruct}), and as such can be a very useful tool in current cosmological research.

The section by section contents of the paper are described as follows: we introduce in section \ref{sphercoords} the Szekeres models in spatial spherical coordinates and in terms of the q--scalars and their exact fluctuations \cite{sussbol}. The Szekeres dipole and its angular extrema are discussed in section \ref{Szdip} (particular cases of Szekeres models from special orientation of the dipole are displayed in Table 1). The notion of the ``LT seed model'' is introduced in section \ref{LTseed}. The conditions that define the location of the spatial extrema of all Szekeres scalars are listed and discussed in section \ref{locextr}. In section \ref{PLH} we introduce Periodic Local Homogeneity, showing that this property is a sufficient condition for the existence of an arbitrary number of spatial extrema of Szekeres scalars. In section \ref{noPLH} we show how spatial extrema can also arise in generic models not compatible with this property by means of a ``simulated shell crossings'' mechanism. The classification of all spatial extrema as maxima, minima and saddle points is found by qualitative arguments in section \ref{classi}. The possibility of avoiding shell crossing singularities in the time evolution of these extrema and possible concavity ``inversions'' of the extrema (a local maximum evolving into a local minimum) are discussed in sections \ref{shx} and \ref{inversion}. The design of Szekeres models that allow for the description of multiple evolving cosmic structures (defined by the spatial maxima and minima of the density) is discussed in section \ref{structures} and a numerical example is examined in section \ref{numex}. A summary and our conclusions are given in section \ref{final}. The article contains four appendices: Appendix \ref{usual} provides the relation between the metric variables and the spherical coordinates we have used and the standard ones, Appendix \ref{regularity} lists the regularity conditions at the origin and to avoid shell crossings, Appendix \ref{extrA} provides a rigorous classification of the spatial extrema and Appendix \ref{formalproofs} the proof of a necessary condition for the existence of spatial extrema of the scalars.   

\section{Szekeres models in spatial spherical coordinates.}\label{sphercoords}

The metric of Szekeres models in spherical coordinates is given by \cite{Bsz2}:
\footnote{By  ``Szekeres models'' we will refer henceforth to quasi--spherical models of class I with a dust source (see details on the obtention and classification of these models in \cite{kras1,kras2,BKHC2009}). The spherical coordinates that we use are those defined as a  ``stereographic'' projection in  \cite{kras2}. The standard diagonal representation of the Szekeres metric and the transformations relating it to (\ref{g1})--(\ref{g3}) are given in Appendix  \ref{usual}.}  
\ba  
g_{tt}&=&-1,\quad g_{rr}=  a^2\Bigg\{ \frac{(\Gamma-\bW)^2}{1-\KK_{q0}r^2}+ \nonumber \\
&{}&\quad +\frac{\sin^4\theta}{(1+\cos\theta)^2}\left[\W^2-2\frac{1+\cos\theta}{\sin^2\theta}\,Z\,\bW\right]\Bigg\}, \label{g1}\\ 
g_{r\theta}&=&\frac{a^2\,r\,\sin\theta}{1+\cos\theta}\left(\bW-Z\right),\; g_{r\phi}=-\frac{a^2\,r\,\sin^2\theta}{1+\cos\theta}\,\bW_{,\phi},\nonumber\\
\label{g2}\\
g_{\theta\theta} &=& a^2 r^2,\quad g_{\phi\phi} = a^2 r^2 \sin^2\theta ,\label{g3}
\ea
where $\KK_{q0}=\KK_{q0}(r)$ is defined in (\ref{qscals}), the time dependence is contained in the scale factors:
\begin{equation} a=a(t,r),\qquad \Gamma=\Gamma(t,r)=1+\frac{ra'}{a},\qquad a'=\frac{\partial a}{\partial r}\label{aGdef},\end{equation}
while the Szekeres dipole $\bW$ and its magnitude $\W$ are given by 
\ba \bW = -X\,\sin\theta\,\cos\phi- Y\,\sin\theta\,\sin\phi-Z\,\cos\theta,\label{dipole}\\
\W^2 \equiv X^2+Y^2+Z^2=\frac{3}{4\pi}\int\limits_{0}^{2\pi}\!{\int\limits_{0}^\pi{\bW^2\, \sin\theta\,\dd\theta\,\dd\phi}},\label{WW} \ea 
where $X=X(r),\,Y=Y(r),\,Z=Z(r)$ are arbitrary functions satisfying the regularity conditions $X(0)=Y(0)=Z(0)=0$ and $X'(0)=Y'(0)=Z'(0)=0$. 

The main covariant scalars associated with Szekeres models are: the density $\rho$; the Hubble scalar $\HH\equiv \Theta/3$ with $\Theta=\nabla_a u^a$; the spatial curvature scalar $\KK=(1/6)\RR$ with $\RR$ the Ricci scalar of the hypersurfaces arbitrary constant $t$;  the eigenvalues $\Sigma,\,\Psi_2$ of the shear and electric Weyl tensors ($\Psi_2$ is the nonzero conformal invariant of Petrov type D spacetimes). These scalars are expressible in the following concise and elegant form:   
\ba  A &=& A_q + \DDa, \qquad A = \rho,\,\HH,\,\KK,\label{Adef}\\
 \Sigma &=& -\DDh,\quad \Psi_2=\frac{4\pi}{3}\DDrho,\label{SigE}\ea
where we have introduced the ``q--scalars'' $A_q$ and their exact fluctuations $\DDa$ (for $A=\rho,\,\HH,\,\KK,$) defined as \cite{sussbol}:
\footnote{The q--scalars and their fluctuations have been widely used in various applications of LT models \cite{sussmodes,part1,part2,perts,RadProfs}. Their properties are extensively discussed in these references. In references \cite{WH2012,Vrba} the density q--scalar $\rho_q$ is denoted by ``$\rho_{AV}$''.} 
\ba  A_q =  \frac{ \int\limits_{r}\!{ \int\limits_{\theta}\!{\int\limits_{\phi}{A\,\FF\sqrt{\JJ}\,\dd r\,\dd \theta\,\dd \phi}}}  }{  \int\limits_{r}\!{ \int\limits_{\theta}\!{\int\limits_{\phi}{\FF\sqrt{\JJ}\,\dd r\,\dd \theta\,\dd y\phi}}}      },\\ \DDa = A-A_q=\frac{r\,A'_q}{3(\Gamma-\bW)},\label{AqDDa}
\ea
where $\FF\equiv \sqrt{1-\KK_{q0}\,r^2}$ and $\JJ$ is the determinant of the spatial part of the metric (\ref{g1})--(\ref{g3}). The q--scalars are related to proper volume averages of their corresponding scalars with weight factor $\FF$ \cite{BoSu2011,sussbol}. At each 2--sphere of constant $r$ the $\DDa$ are exact fluctuations around this average, which defines an FLRW background as $r\to\infty$. The q--scalars and their fluctuations are covariant objects \cite{sussbol} and reduce in their linear limit to standard variables of cosmological perturbations in the isochronous comoving gauge \cite{part2,perts}. 

Evaluating the integral in (\ref{AqDDa}) for each scalar $A$ yields the following scaling laws and constraints
\ba \rho_q &=& \frac{\rho_{q0}}{a^3},\qquad \KK_q=\frac{\KK_{q0}}{a^2},\qquad \HH_q=\frac{\dot a}{a},\label{qscals}\\
 \HH_q^2 &=& \frac{\dot a^2}{a^2}=\frac{8\pi}{3}\rho_q -\KK_q+\frac{8\pi}{3}\Lambda,\label{Friedeq}\ea
 \ba 2\HH_q\,\DDh &=& \frac{8\pi}{3}\DDrho-\DDKK,\label{DDslaws1}\\
 \frac{\DDrho}{\rho_q}-\frac{3}{2}\frac{\DDKK}{\KK_q} &=&\frac{1-\bW}{\Gamma-\bW}\left[\frac{\DDrho_0}{\rho_{q0}}-\frac{3}{2}\frac{\DDKK_0}{\KK_{q0}}\right],\label{DDslaws2}\ea
where the subindex ${}_0$ denotes (and will denote henceforth) evaluation at an arbitrary fixed $t=t_0$, thus allowing for the study of the dynamics of the models within an initial value framework.  Every Szekeres model becomes fully specified by initial conditions furnished by six free parameters: $\Lambda$,\,two of $\{\rho_{q0},\,\HH_{q0},\,\KK_{q0}\}$ plus the dipole parameters $\{X,\,Y,\,Z\}$. The models become determined either by solving the Friedman equation (\ref{Friedeq}) or by integrating the evolution equations for the $A_q,\,\DDa$ (see  \cite{sussbol,nuevo}).

\begin{figure}
\begin{center}
\includegraphics[scale=0.3]{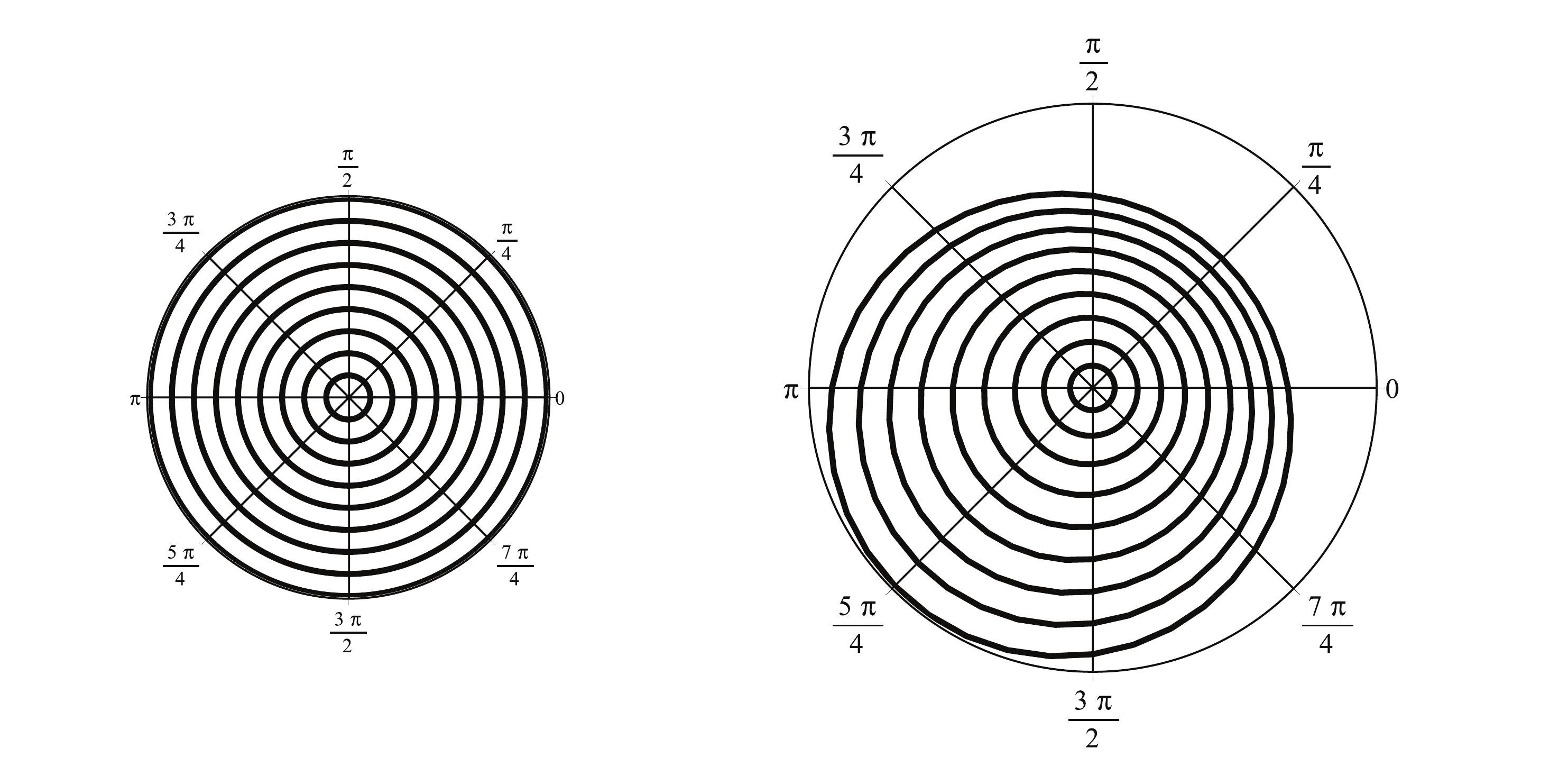}
\caption{{\bf Foliation by non--concentric 2--spheres.} Every time slice of a Szekeres model is smoothly foliated by non--concentric spheres marked by arbitrary constant $r$ with surface area $4\pi a^2 r^2$. The right hand side figure depicts the equatorial projection of the level curves of constant proper length (along radial rays) of the 2--spheres of constant $r$ in the coordinate diagram in the left hand side figure. Plotted in terms of proper length illustrates how the 2--spheres are not orthogonal to the radial rays, which gives rise to the non--diagonal metric components in (\ref{g1})--(\ref{g2}). The difference between the panels of this figure is crucial to understand that the apparent rotational symmetry (as in the left panel) seen in figures \ref{AEcurves}, \ref{LocHomSph}, \ref{location}, \ref{density} and \ref{pancake} is a coordinate effect without covariant meaning: if these figures were plotted with respect to proper radial distances their morphology  would look like the right panel. }
\label{twospheres}
\end{center}
\end{figure}

It is very important to emphasise that the spherical coordinates we are using lack a covariant meaning. Hence, ``angular directions'' indicate locations along curves with constant $(\theta,\phi)$ (radial rays), which are not spacelike geodesics as in spherically symmetric spacetimes \cite{RadAs}. The origin worldline 
\footnote{We assume henceforth that the LT seed model (section \ref{LTseed}) admits a symmetry centre at $r=0$, and thus its associated Szekeres models admit an origin worldline whose regularity conditions are given in Appendix \ref{origin}. The symmetry centre or origin worldline is a physically motivated assumption, but it is not absolutely necessary, as regular LT and Szekeres and models exist that admit either two such worldlines or none (for time slices with the topology of a 3--sphere or of a ``wormhole'' \cite{szwh}). The results we obtain in the present article can be easily extended to this type of models (see Appendix D of \cite{sussbol}). }
is not a symmetry centre. However, as opposed to the standard dipole coordinates $(p, q$) (see Appendix \ref{usual}), the angular coordinates $(\theta,\phi)$ are bounded and thus are very useful to specify coordinate locations and for an intuitive understanding of the properties of the models, for example: it is evident from the metric (\ref{g1})--(\ref{g3}) that the surfaces of constant $t$ and constant $r$ are 2--spheres, as setting $\dd t=\dd r=0$ yields the metric of a 2--sphere with surface areas $4\pi a^2 r^2$. These surfaces constitute a smooth foliation of any time slice by non--concentric 2--spheres (see figure \ref{twospheres}). This follows from the fact that $g_{rr}$ depends on the angular coordinates, hence the proper radial length along radial rays from $r=0$ 
to points in any 2--sphere $\ell=\int_0^r{\sqrt{g_{rr}\dd r}}$ smoothly depends on the angular coordinates of the points. 

\section{Angular orientation: the Szekeres dipole.}\label{Szdip}
 
It follows directly from (\ref{g1})--(\ref{g3}), (\ref{Adef}), (\ref{SigE}) and (\ref{AqDDa}) that the angular coordinate dependence of the metric and of all covariant scalars is entirely contained in the Szekeres dipole $\bW$ \cite{kras2}. The spherical coordinates allow us to obtain a very precise  specification of a unique angular coordinate location of the dipole at each 2--sphere of constant $r$ from the extrema of each 2--sphere: the ``{\bf angular extrema}'' that follow from the condition: 
\begin{equation} \bW_{,\theta}=\bW_{,\phi}=0,\label{angextW}\end{equation}
which yields two antipodal positions in the $(r,\theta,\phi)$ coordinates:
\ba    \phi_{-} &=& \arctan \left(\frac{Y}{X}\right),\quad \theta_{-}= \arccos\left(\frac{Z}{\sqrt{X^2+Y^2+Z^2}}\right)\nonumber\\
\label{sol1}\\
  \phi_{+} &=& \pi+\phi_{-},\qquad \theta_{+}= \pi-\theta_{-}.\label{sol2}\ea 
Since $r$ varies smoothly along the 2--spheres, the solutions (\ref{sol1})--(\ref{sol2}) define in the spherical coordinate system the following two curves parametrised by $r$ (see figure \ref{AEcurves}):
\begin{figure}
\begin{center}
\includegraphics[scale=0.4]{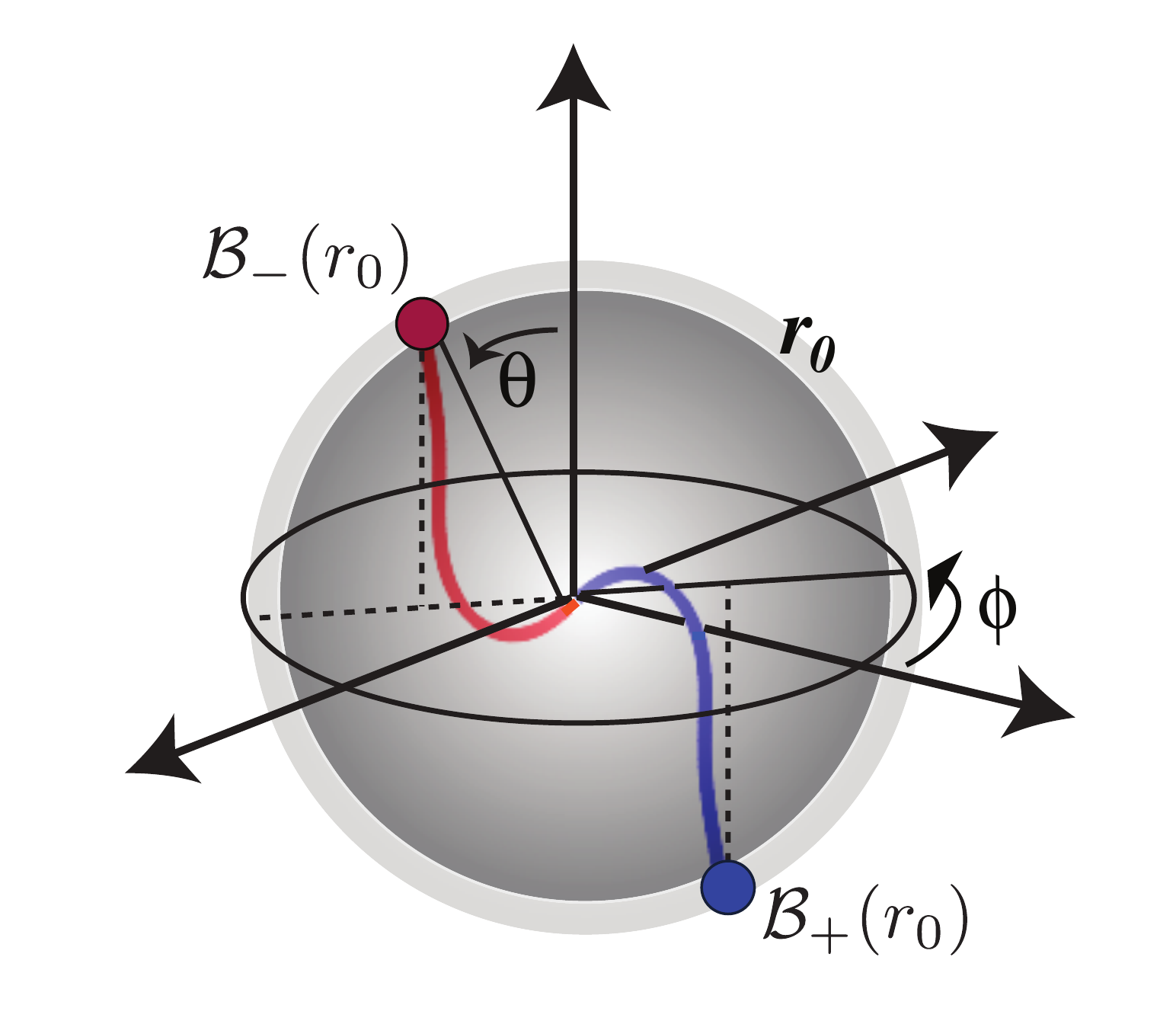}
\caption{{\bf Curves of angular extrema.} The curves $\B_\pm(r)$ in the coordinate system $(r,\theta,\phi)$ defined in (\ref{C12}) that follow from the angular extrema condition (\ref{angextW}) are depicted in blue ($\B_{+}(r)$) and red ($\B_{-}(r)$) (colours only appear in the online version). These curves determine for every 2--sphere marked by arbitrary constant $r=r_0$ two points with coordinates $\B_{+}(r_0)$ and $\B_{-}(r_0)$ where the Szekeres dipole has, respectively, an angular maximum and an angular minimum. These two points define from (\ref{sol1})--(\ref{sol2}) two antipodal angular directions $[\theta_{+}(r_0),\phi_{+}(r_0)]$ and $[\theta_{-}(r_0),\phi_{-}(r_0)]$ associated with the angular extrema at each $r=r_0$. The full spatial extrema of all Szekeres scalars necessarily lie in points along the curves $\B_\pm(r)$ at values of $r$ that shift in time. Particular cases of Szekeres models occur when the curves $\B_\pm(r)$ are constrained to a radial ray or a plane in $(r,\theta,\phi)$. See Table 1. If the dipole parameters are defined as piecewise functions (as in (\ref{pwXYZ})), the curves $\B_\pm(r)$ become piecewise segments (see figures \ref{location} and \ref{density}).}
\label{AEcurves}
\end{center}
\end{figure}
\begin{equation}  \B_{\pm}(r )=[r,\,\theta_{\pm}(r ),\,\phi_{\pm}( r)], \label{C12}\end{equation}
which cross the origin. The magnitude of $\bW$ at the curves $\B_\pm(r)$ is 
\begin{equation}\bW_{\pm}\equiv \bW(r,\theta_\pm(r),\phi_\pm(r))=\pm\,\W(r),\label{WAE}\end{equation}
where the subindex ${}_\pm$ denotes (and will denote henceforth)  evaluation along the curves $\B_\pm(r)$ and $\W$ is the dipole magnitude defined in (\ref{WW}). Therefore, the angular extrema at each surface of constant $r$ necessarily lie in their intersection with the curves $\B_\pm(r)$ (see figure \ref{AEcurves}). Since $\bW>0$ along $\B_{+}(r)$ and $\bW<0$ along $\B_{-}(r)$ we have at each 2--sphere of constant $r$
\ba  \hbox{Angular maxima of}\,\, \bW\,\, \hbox{lie along}\,\, \B_{+}(r),\label{WAEmax}\\
 \hbox{Angular minima of}\,\, \bW\,\, \hbox{lie along}\,\, \B_{-}(r),\label{WAEmin}\\
 \hbox{Saddle point of}\,\, \bW\,\, \hbox{at}\,\, r=0,\label{WAEsaddle}\ea
while allowing for the radial dependence of $\bW$ the full spatial 3--dimensional extrema of $\bW$ are located in the curves $\B_\pm(r)$ at some $r=r_e$ such that $XX'+YY'+ZZ'=0\,\,\Rightarrow\,\,\W'=0$. 

Restrictions on the orientation of the dipole in the coordinates $(r,\theta,\phi)$ correspond to particular cases of Szekeres models that follow from restrictions on the parameters $X,\,Y,\,Z$. These particular cases are listed in Table 1.
\begin{table*}
\begin{center}
\begin{tabular}{|c| c| c|}
%\hline
%\hline
%\hline
%\hline
%\hline
\hline
\multicolumn{3}{|c|}{{\bf Dipole oriented along a radial ray: one free parameter.}}
\\  
\hline
\hline
{Parameter restrictions} &{Spherical coordinates} &{Rectangular coordinates} 
\\
\hline
\hline
{$X\ne 0,\,\,Y=Z=0$} &{$\theta_\pm=\frac{\pi}{2},\,\phi_\pm=0,\pi$} &{$[x,0,0]$}
\\  
\hline
%\hline
{$Y\ne 0,\,\,X=Z=0$} &{$\theta_\pm=\frac{\pi}{2},\,\phi_\pm=\frac{\pi}{2},\frac{3}{2}\pi$} &{$[0,y,0]$}
\\
\hline  
{$Z\ne 0,\,\,X=Y=0$} &{$\theta_\pm=0,\pi$} &{$[0,0,z]$} 
\\
\hline  
{$X = a_0f,\,Y=b_0f,\,Z=c_0f$} &{$\tan\phi_\pm=\pm(b_0/a_0)$} &{$[a_0x,b_0y,c_0z]$}
\\ 
{two of $a_0,b_0,c_0$ nonzero, $f=f(r)$} &{$\tan\theta_\pm=\pm\sqrt{a_0^2+b_0^2}/c_0$} &{(general ray)}
\\
%%%%%----
%%%%%%---- 
%\hline
%\hline
%\hline
%\hline
\hline
\hline
%%%%%------
%%%%%------

%%%%%%------
%%%%%%------
\multicolumn{3}{|c|}{{\bf Dipole contained in a plane crossing the origin: two free parameters.}}
\\  
\hline
\hline
{Parameter restrictions} &{Spherical coordinates} &{Rectangular coordinates} 
\\
\hline
\hline
{$X,\,Y\ne 0,\,\,Z=0$} &{$\theta_\pm=\frac{\pi}{2}$} &{$[x,y,0]$} 
\\
\hline
{$Z,\,Y\ne 0,\,\,X=0$} &{$\phi_\pm=0,\pi$} &{$[0,y,z]$} 
\\
\hline
{$Z,\,X\ne 0,\,\,Y=0$} &{$\phi_\pm=\frac{\pi}{2},\frac{3\pi}{2}$} &{$[x,0,z]$} 
\\
\hline
{$a_0X+b_0Y+c_0Z=0$} &{$b_0\cot\theta_\pm=$} &{$a_0x+b_0y+c_0z=0$} 
\\
{$a_0,b_0,c_0$ nonzero} &{$-a_0\cos\phi_\pm-b_0\sin\phi_\pm$} &{(general plane)}
\\
%\hline 
%\hline
%\hline
%\hline
%\hline
\hline
%%%%-------
%%%%-------

%%%%-------
%%%%-------
\end{tabular}
\end{center}
\caption{{\bf{Particular cases vs dipole orientation}}. The angular extrema (\ref{sol1})--(\ref{sol2}) determine the orientation of the Szekeres dipole $\bW$ given by (\ref{dipole}) at every 2--sphere of constant $r$. The table displays the dipole parameters $X,\,Y,\,Z$ (left column) that follow from restricting this orientation (i) to lie along a radial ray and (ii) to be contained in a plane crossing the origin. All cases in (i) are equivalent and can be related by rotations in the coordinate system $(r,\theta,\phi)$ (the same remark applies to the cases (ii)). The cases in (i) seem to correspond to axially symmetric Szekeres models (see section 3.3.1 of \cite{BKHC2009}). The dipole orientation is given in spherical coordinates (middle column) and in rectangular coordinates (right column) obtained by the standard transformations $x=r\sin\theta\cos\phi,\,y=r\sin\theta\sin\phi,\,z=r\cos\theta$. For these particular cases the spatial extrema of Szekeres scalars must lie in their corresponding directions.}
\label{tabla1}
\end{table*}
%%%

\section{The LT seed model.}\label{LTseed}  

We shall use the term  ``{\bf LT seed model}'' to denote the unique spherically symmetric LT model that follows by setting $\bW=0$ in the metric (\ref{g1})--(\ref{g3}) and in the covariant scalars (\ref{Adef})--(\ref{SigE}). Evidently,  every Szekeres model can be constructed from its LT seed model by specifying the dipole parameters $X,\,Y,\,Z$ that define $\bW$. Therefore, Szekeres models inherit some of the properties of their LT seed model: the scale factors $a,\,\Gamma$ and the q--scalars in (\ref{qscals}) are common to both. In fact, the volume integral (\ref{AqDDa}) evaluated for LT scalars yields the same q--scalars and Friedman equation  (\ref{qscals}) and (\ref{Friedeq}) \cite{sussmodes,part1,perts}. Hence, the LT scalars satisfy the same relations (\ref{Adef})--(\ref{SigE})
\footnote{Notice that the density LT fluctuation $\DDrholtb$ is exactly the term ``$\rho_{LT}-\rho_{AV}$'' that frequently appears in \cite{WH2012,Vrba}.}
:
\ba \Altb=A_q+\DDaltb,\quad \Sigltb=-\DDhltb,\quad \Pltb=\frac{4\pi}{3}\DDrholtb,\nonumber\\
\label{LTscals}\ea 
with the exact LT fluctuations $\DDaltb$ related to Szekeres fluctuations $\DDa$ by:
\ba \DDaltb = \Altb-A_q =\frac{r\,A'_q}{3\Gamma}\nonumber\\ 
\Rightarrow\quad \DDa = \frac{\Gamma}{\Gamma-\bW}\,\DDaltb.
\label{DDaltb}\ea
Since the LT fluctuations $\DDaltb$ only depend on $(t,r)$, it is useful to express the Szekeres scalars $A=\rho,\,\HH,\,\KK$ and $\Sigma,\,\Psi_2$ exclusively in terms of LT objects and $\bW$, either as: 
\begin{equation}  A = A_q + \frac{\Gamma\,\DDaltb}{\Gamma-\bW},\quad \Sigma = -\frac{\Gamma\,\DDhltb}{\Gamma-\bW},\quad \Psi_2 = \frac{4\pi}{3}\,\frac{\Gamma\,\DDrholtb}{\Gamma-\bW}, \label{ASigPsi1}\end{equation}
or, alternatively, in terms of $\Altb$ and $\Sigltb,\,\Pltb$
\begin{equation} A = \Altb+ \frac{\DDaltb\,\bW}{\Gamma-\bW},\qquad \Sigma = \frac{\Gamma\,\Sigltb}{\Gamma-\bW},\qquad \Psi_2 = \frac{\Gamma\,\Pltb}{\Gamma-\bW}. \label{ASigPsi2}\end{equation}
with the resulting advantage in both decompositions that all the angular dependence is contained in the dipole term $\bW$. However, the decomposition (\ref{ASigPsi1}) is more useful, and is thus preferable, because the q--scalars satisfy simple scaling laws like (\ref{qscals}) and thus are easier to manipulate than the standard LT scalars $\Altb$. We assume henceforth that regularity conditions to avoid shell crossings (\ref{noshx1})--(\ref{noshx3}) hold everywhere, hence 
\begin{equation} \DDa=0\quad\Leftrightarrow\quad \DDaltb=0,\quad \hbox{sign of}\,\,\DDa = \hbox{sign of}\,\,\DDaltb\label{SzLTflucts}\end{equation}
hold everywhere and thus we can consider the zeroes and signs of both fluctuations interchangeably.  

\section{Location of the spatial extrema of Szekeres scalars.}\label{locextr}

The conditions for the existence of a spatial 3--dimensional local extremum of $A=\rho,\,\HH,\,\KK$ at any time slice are given by
\begin{equation} A'=0,\qquad A_{,\theta}=0,\qquad A_{,\phi}=0,\label{condAextr}\end{equation}
with similar conditions for $\Sigma$ and $\Psi_2$ (unless stated otherwise all  extrema defined by (\ref{condAextr}) will be local). We examine separately below the angular and radial derivatives. 

\subsection{Angular extrema.}\label{AEofA} 

We obtain from (\ref{ASigPsi1}) the conditions for the angular extrema of $A=\rho,\,\HH,\,\KK$ at any fixed $t$:
\ba  A_{,\theta}=\frac{\DDa\,\bW_{,\theta}}{\Gamma-\bW}=\frac{\DDaltb\,\Gamma\,\bW_{,\theta}}{(\Gamma-\bW)^2}=\frac{r\,A'_q\,\bW_{,\theta}}{3(\Gamma-\bW)^2}=0,\nonumber\\
 A_{,\phi}=\frac{\DDa\,\bW_{,\phi}}{\Gamma-\bW}=\frac{\DDaltb\,\Gamma\,\bW_{,\phi}}{(\Gamma-\bW)^2}=\frac{r\,A'_q\,\bW_{,\phi}}{3(\Gamma-\bW)^2}=0,\nonumber\\
 \label{AEab}\ea
with analogous expressions for $\Sigma$ and $\Psi_2$. Evidently we have $\bW_{,\theta}=\bW_{,\phi} =0\,\,\Rightarrow\,\,A_{,\theta}=A_{,\phi}=0$, but the converse is false. However, the condition $\DDa=0$ defines for all $(\theta,\phi)$ a trivial or {\it degenerate} angular extremum of $A$ 
\footnote{The condition $\DDa=0$ may hold in specific values $(t,r)$ for all $(\theta,\phi)$ (see section \ref{PLH}). It implies $A=A_q=\Altb$ (from (\ref{AqDDa}) and (\ref{DDaltb})), and thus $A$ becomes independent of the angular coordinates. The determinants of the Hessian matrix and all its minors (see \ref{extrA}) vanish at points that fulfil $\DDa=0$, and thus this condition marks the same {\it degenerate} angular extrema of $A$ as a constant function for which all derivatives are trivially zero. The equality $A=A_q=\Altb$ also holds at coordinate values such that $\bW(r,\theta,\phi)=0$, but the derivatives of $\bW$ do not vanish at these values.}
. As we explain in section \ref{PLH}, this condition leads to the Comoving Homogeneity Spheres. Therefore, we have the following important result:   
\begin{quote}
The (non--degenerate) angular extrema of all Szekeres scalars $A=\rho,\,\HH,\,\KK$, as well as $\Sigma$ and $\Psi_2$, coincide at all $t$ with the angular extrema of the Szekeres dipole $\bW$. Since  $A_{,\theta}=A_{,\phi}=0$ is a necessary (not sufficient) condition for (\ref{condAextr}), the spatial extrema of all Szekeres scalars are necessarily located along the curves $\B_\pm(r)$.      
\end{quote}
\subsection{Radial location of the spatial extrema: extrema of the radial profiles.} \label{perfiles}

The missing condition in (\ref{condAextr}) for a spatial extremum of $A=\rho,\,\HH,\,\KK$ is the radial equation $A'=0$ for $A$ from (\ref{ASigPsi1}) (and similar equations $\Sigma'=0$ and $\Psi_2'=0$ from  (\ref{ASigPsi2})). However, the spatial extrema are necessarily located in the curves $\B_\pm(r)$, hence these radial conditions must be evaluated for the ``radial profiles'' of the scalars evaluated in these curves (which for fixed $t$ depend only on $r$). As a consequence, finding the location of the spatial extrema reduces to finding the extrema of the  ``radial profiles'' of the scalars given explicitly by:
\ba  A_\pm =A_q + \frac{\Gamma\,\DDaltb}{\Gamma\mp\W} = \Altb \pm \frac{\W\,\DDaltb}{\Gamma\mp\W}\label{ASigPsi3a}\\
  \Sigma_\pm = -\frac{\Gamma\,\DDhltb}{\Gamma\mp\W},\qquad \Pspm = \frac{4\pi}{3}\,\frac{\Gamma\,\DDrholtb}{\Gamma\mp\W}, \label{ASigPsi3b}\ea
where $\W$ and $\DDaltb$ are defined in (\ref{WW}) and (\ref{DDaltb}), and we have used (\ref{WAE}) and (\ref{ASigPsi1})--(\ref{ASigPsi2}).  %
The radial conditions become for $A_{+}$ and $A_{-}$
\ba A'_{+}=0\,\, \Rightarrow\,\, 
\left[1+3(\Gamma-\W)-\frac{r(\Gamma'-\W')}{\Gamma-\W}\right]A'_q+rA''_q=0,\nonumber\\
\label{condr1}\\
 A'_{-}=0\,\, \Rightarrow\,\, 
\left[1+3(\Gamma+\W)-\frac{r(\Gamma'+\W')}{\Gamma+\W}\right]A'_q+rA''_q=0,\nonumber\\\label{condr2}\ea
and for $\Sigma_\pm$ and $\Pspm$
\ba  \Sigma'_\pm=0\quad \Rightarrow\quad \left[1-\frac{r(\Gamma'\mp\W')}{\Gamma\mp\W}\right]\,\HH'_q+r\HH''_q=0,\nonumber\\
  \Pspm'=0\quad \Rightarrow\quad \left[1-\frac{r(\Gamma'\mp\W')}{\Gamma\mp\W}\right]\,\rho'_q+r\rho''_q=0.\nonumber\\
  \label{condrSP}\ea
where we used (\ref{DDaltb}) to eliminate $\DDaltb$ and $[\DDaltb]'$ in terms of $A'_q$ and $A''_q$. The solutions of (\ref{condr1})--(\ref{condr2}) or (\ref{condrSP}) yield the extrema of the radial profiles (\ref{ASigPsi3a})--(\ref{ASigPsi3b}), which together with the angular extrema satisfying (\ref{AEab}) lead to the full 3--dimensional spatial extrema complying with (\ref{condAextr}). Therefore, the location of all spatial extrema of Szekeres scalars in the spherical coordinates is given by
\begin{equation} \left[\, t_{e\pm},\,r_{e\pm},\,\theta_\pm(r_{ e\pm}),\,\phi_\pm(r_{e\pm})\,\right],\label{extrcoords}\end{equation}
where $\theta_\pm,\,\phi_\pm$ are given by (\ref{sol1})--(\ref{sol2}) and $(t_{e\pm},r_{e\pm})$ are solutions of the radial conditions (\ref{condr1})--(\ref{condr2}) (for extrema of $A$) or (\ref{condrSP}) (for extrema of $\Sigma$ and $\Psi_2$). We can ascertain the following points before finding (or discussing existence) of solutions of the radial conditions:  
\begin{itemize}
\item {\bf The regular origin is a spatial extremum for every scalar.} If standard regularity conditions hold (see \ref{origin}), (\ref{AEab}) and (\ref{condr1})--(\ref{condr2}) as well as (\ref{condrSP}) (and thus (\ref{condAextr})) hold at $r=0$ for all $t$.
\item {\bf The spatial extrema of all scalars are not comoving.} The constraints (\ref{condr1})--(\ref{condr2}) and (\ref{condrSP}) depend on time, hence their solutions are (in general) different for different values of $t$ and define constraints of the form $r_{e\pm}=r_{e\pm}(t)$. However, to simplify the notation we will denote these radial coordinates simply by $r_{e\pm}$.   
\item {\bf Spatial extrema of different scalars.} Evidently, the radial conditions (\ref{condr1})--(\ref{condr2}) have different solutions for different scalars.   
\end{itemize}
However, the location and classification of these extrema cannot be achieved as long as  the main technical difficulty remains: finding solution of the radial conditions (\ref{condr1})--(\ref{condr2}) and (\ref{condrSP}) for $r>0$. Evidently, solving these radial conditions, for $r>0$ and for generic models and without further assumptions, is practically impossible without numerical methods. However, it is possible to obtain sufficient conditions for the existence of such solutions in specified comoving radial ranges without actually having to solve these constraints.

\section{Sufficient conditions for the existence of spatial extrema.}\label{PLH} 

\begin{figure}
\begin{center}
\includegraphics[scale=0.35]{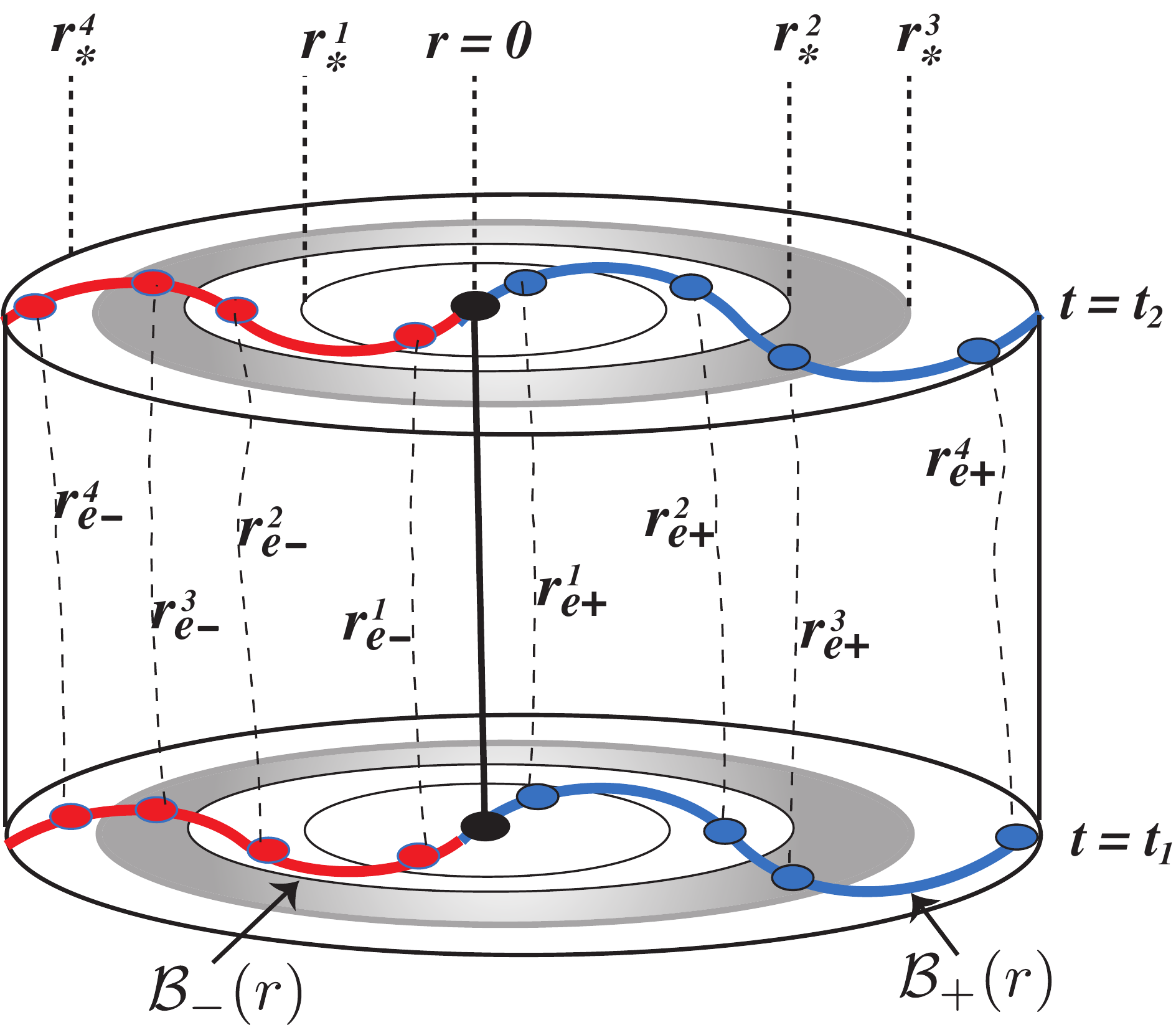}
\caption{{\bf Spacetime diagram of a Szekeres model admitting Peridic Local Homogeneity (PLH).} A qualitative schematic view of the spacetime evolution of four Comoving Homogeneity Spheres represented as circles marked by radial coordinates $r_*^i$ that follow from assuming local homogeneity conditions (\ref{localhom}) that define PLH. The extremum at the origin (spatial maximum or minimum) is marked by a thick black dot, while four pairs of spatial extrema of a Szekeres scalar are marked as blue or red thick dots, depending on their location along one of the curves of angular extrema $\B_{\pm}(r)$ depicted by blue and red curves (colours only appear in the online version). As we show in section \ref{classi} (see also Appendix \ref{extrA}) , the extrema along $\B_{-}$ must be saddle points, while those along $\B_{+}$ can be spatial maxima, minima or saddles. The origin worldline is depicted as a solid black line, whereas the dashed curves represent the non--comoving wordlines of the four pairs of spatial extrema with spatial coordinates $[r_{e\pm}^i,\,\theta_{\pm}(r_{e\pm}^i),\,\phi_{\pm}(r_{e\pm}^i)]$ with  $\theta_{\pm},\phi_{\pm}$ given by (\ref{sol1})--(\ref{sol2}). The shaded area represents one of the comoving shell regions $[t,\Delta_*^i,\theta,\phi]$ between the Comoving  Homogeneity Spheres where the extrema evolve for all $t$ (pending absence of shell crossings, see section \ref{shx}). The curves $\B_\pm(r)$ become piecewise continuous segments (thus providing a more precise angular location of the extrema) for dipole parameters given as in (\ref{pwXYZ}) (see figures \ref{location} and \ref{density})}
\label{LocHomSph}
\end{center}
\end{figure}

Sufficient conditions for the existence of spatial extrema of Szekeres scalars follows from imposing ``local homogeneity'' defined (in a covariant manner) by demanding that the shear and Weyl tensors vanish (which implies $\Sigma=\Psi_2=0$) for all $t$, but only for selected worldlines or surfaces (as opposed to global homogeneity leading to the FLRW limit if this condition holds everywhere \cite{kras1,kras2,sussbol}). In fact, such local homogeneity holds at the origin worldline $r=0$    
if standard regularity conditions are satisfied (see \ref{origin}). This type of local homogeneity can also be imposed on a discrete set of comoving 2--spheres marked by fixed radial comoving coordinate values $r>0$:

\begin{quote}
{\bf Periodic Local Homogeneity (PLH).} Consider a sequence of $n$ arbitrary nonzero increasing radial comoving coordinate values and $n$ open intervals between them
\ba  r_*^i &=& r_*^1,\,..,\,r_*^n,\nonumber\\
 \Delta_*^i &=& 0<r<r_*^1,\,..\,,r_*^{n-1}<r<r_*^n.\nonumber\\
 \label{rstar}\ea
A Szekeres model is compatible with PLH in the spacetime region bounded by the spherical world--tube  $(t,r_*^n,\theta,\phi)$ (see figure \ref{LocHomSph}) if at each 2--sphere $(r_*^i,\theta,\phi)$ the scalars $A=\rho,\,\HH,\,\KK$  comply at all $t$ with 
\ba \qquad \Sigma_* &=& [\Psi_2]_*=0 \nonumber\\ 
\quad &\Rightarrow&\quad
[\DDa]_{*} = [\DDaltb]_*=[A'_q]_*=0,\nonumber\\
\quad &\Rightarrow&\quad A_* = A_{q*}=[\Altb]_*,\nonumber\\
\label{localhom}\ea
where the subindex ${}_*$ denotes evaluation at the $n$ fixed comoving radii (\ref{rstar}). We will use the term {\bf Comoving Homogeneity Spheres} to denote the non--concentric 2--spheres $(r_*^i,\theta,\phi)$ where local homogeneity holds. The origin worldline  $r=0$ can be regarded as the Comoving Homogeneity Spheres of zero area.
\end{quote}
Szekeres models admitting PLH are characterised by the following properties:
\begin{itemize}
\item They can be specified by initial conditions in which any two of the tree initial value functions $A_{q0}=\rho_{q0},\,\KK_{q0},\,\HH_{q0}$ comply with
\ba &{}& A'_{q0}(r_*^i) = 0\nonumber\\
&{}& \Rightarrow\,\, 
[\DDaltb]_0(r_*^i) = 0\,\,\Rightarrow\,\,\DDa_0(r_*^i,\theta,\phi)=0,\nonumber\\
&{}& \Rightarrow\,\, 
A_0(r_*^i)  A_{q}(r_*^i)=[\Altb]_0(r_*^i),\nonumber\\ 
&{}& \Rightarrow\,\,\Sigma_0(r_*^i)=[\Psi_2]_0(r_*^i)=0.\nonumber\\
\label{localhom1ab}\ea
Notice that these initial conditions are sufficient to define PLH at all $t$ (pending shell crossings), since (\ref{SigE}) and the constraints (\ref{DDslaws1})--(\ref{DDslaws2}) and (\ref{DDaltb}) that relate the $\Sigma,\,\Psi_2$ and the fluctuations $\DDa$ and are preserved for the whole time evolution  
\footnote{If condition $\DDa=\DDaltb=0$ holds only for one of the scalars $A=\rho,\,\HH,\,\KK$, it will not hold for the remaining ones (see examples for generic LT models in \cite{RadProfs}). Since $\Sigma=\Psi_2=0$ does not hold (though one of these scalars may vanish because of (\ref{SigE})), this case does not lead to PLH. Also, $\DDa=0$ for a single scalar does not occur in comoving surfaces and (in general) it only holds for restricted ranges of the time evolution. We discuss this case in section \ref{inversion}.}
.
We only need two of the three scalars to comply with (\ref{localhom}) and (\ref{localhom1ab}) because the fluctuations are related by these constraints.
\item They become partitioned by $n$ inhomogeneity shell regions $(t,\Delta_*^i,\theta,\phi)$ (see figure \ref{LocHomSph}) bounded by  Comoving Homogeneity Spheres. Since condition (\ref{localhom}) that defines these 2--spheres holds in an asymptotic FLRW background \cite{perts}, each pair of these contiguous spheres plays the role of time preserved localised FLRW backgrounds that surround these inhomogeneity shells where the extrema of the scalars are located. 
\end{itemize}

Since the zeroes and signs of the $\DDa$ determine the concavity pattern (maxima and minima) of the radial profiles of all Szekeres scalars (\ref{ASigPsi3a})--(\ref{ASigPsi3b}), we have the following result:
\begin{quote}
{\bf Proposition 1.} PLH defined by (\ref{localhom}) and (\ref{localhom1ab})  constitutes a sufficient condition for the existence of $2n$ spatial extrema (not located in the origin) of the Szekeres scalars $A=\rho,\,\HH,\,\KK$ and of $\Sigma,\,\Psi_2$ for all $t$. 
\end{quote}
Besides the extremum at $r=0$ (see section \ref{locextr}), models compatible with PLH admit $2n$ extrema marked by coordinate values (\ref{extrcoords}) with  radial coordinates $r_{e\pm}^1,..,r_{e\pm}^n$ and distributed in $n$ pairs as follows:
\begin{itemize}
\item  the radial coordinates of each pair $r_{e\pm}^i$ are (for each scalar) solutions of the radial conditions (\ref{condr1})--(\ref{condr2}) or (\ref{condrSP}) in the intervals  $0\leq r_{e\pm}^1\leq r_*^1,..,r_*^{n-1}\leq r_{e\pm}^n\leq r_*^n$, with the $r_*^i$ given by (\ref{rstar}). 
\item spatial extrema at $r_{e+}^i>0$ are located in the curve $\B_{+}(r)$ and those at $r_{e-}^i>0$ in the curve $\B_{-}(r)$.  
\end{itemize}
{\bf Proof}. As we prove in Appendix \ref{formalproofs}, PLH defined by (\ref{localhom}) is a sufficient condition for the existence, at all $t$, of $n$ extrema of each of the radial profiles $A_\pm,\,\Sigma_\pm,\,[\Psi_2]_\pm$ in the radial ranges $0\leq r_{{\rm tv}\pm}^1\leq r_*^1,..,r_*^{n-1}\leq r_{{\rm tv}\pm}^n\leq r_*^n$. Therefore, the radial coordinates $r_{e\pm}^i$ of the $2n$ pairs of spatial extrema with $r>0$ correspond to the radial coordinates $r_{{\rm tv}\pm}^i$ of the extrema with $r>0$ of $A_\pm$ and $\Sigma_\pm,\,[\Psi_2]_\pm$ given in (\ref{ASigPsi3a})--(\ref{ASigPsi3b}), as these are the radial profiles of the scalars $A=\rho,\,\HH,\,\KK$ and $\Sigma,\,\Psi_2$ along the curves $\B_\pm(r)$, and thus  satisfy (\ref{AEab}). Since they are solutions of the radial constraints (\ref{condr1})--(\ref{condr2}) or (\ref{condrSP}) for all $t$, they also satisfy (\ref{condAextr}). Since PLH is preserved by the time evolution, the existence and location of the extrema at the intervals $\Delta_*^i$ in the curves $\B_\pm(r)$ are also preserved in time (pending shell crossings, see section \ref{shx}).    
%   
%LocHomSph3.pdf
%
%
%All4Profiles2.pdf 
%
\begin{figure}
\begin{center}
\includegraphics[scale=0.4]{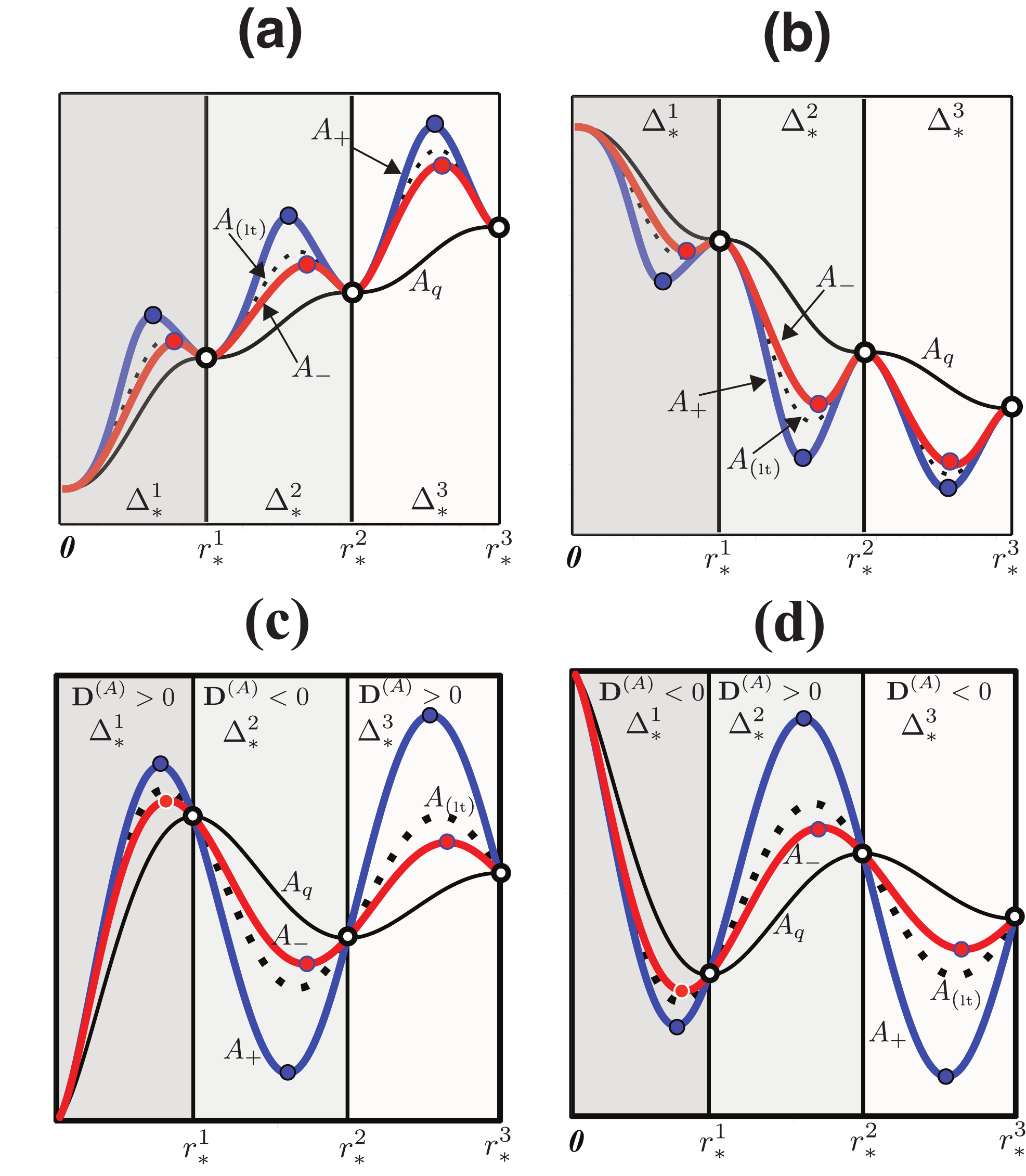}
\caption{{\bf Concavity patterns of radial profiles under the assumption of PLH.} The panels depict four of the possible patterns of the concavity of the radial profiles (\ref{ASigPsi3a}) at an arbitrary $t$ that follow (qualitatively) from the inequalities (\ref{ineq12})--(\ref{ineq34}) applied to the restrictions that arise from PLH. Panels (a) and (b) correspond to $\DDa$ keeping the same sign in the $\Delta_*^i$, while (c) and (d) describe the patterns when this sign alternates. The profiles of $A_q,\,\Altb,\,A_{+},\,A_{-}$ are depicted by the solid black, dashed, blue and red curves (colours only appear in the online version), within the three radial intervals $\Delta_*^1,\,\Delta_*^2,\,\Delta_*^3$ (shaded rectangles) bounded by the comoving coordinates $r_*^i=0,r_*^1,r_*^2,r_*^3$. The extrema of $A_q$ (white dots) coincide with the $r_*^i$, while the extrema of $\Altb$ and $A_\pm$ (marked by red and blue thick dots) form a pattern of maxima, minima or alternating maxima and minima. Similar patterns occur for the radial profiles $\Sigma_\pm$ and $[\Psi_2]_\pm$ given by (\ref{ASigPsi3b}). These profiles are preserved through the full time evolution, pending shell crossings or concavity inversions (see sections \ref{shx} and \ref{inversion}). }
\label{figprofiles}
\end{center}
\end{figure}

Proposition 1 can also be proven by qualitative arguments assisted by figure \ref{figprofiles}.
The following inequalities that follow directly from (\ref{ASigPsi3a})--(\ref{ASigPsi3b}) hold at all $t$ for any Szekeres model complying with standard regularity conditions (see Appendixes \ref{origin} and \ref{noshx})  
\ba
 A_{+}<\Altb<A_q\,\,\hbox{and}\,\,  \Altb < A_{-}<A_q \,\, \hbox{if}\,\,\DDa<0,\nonumber\\
A_{+}>\Altb>A_q\,\,\hbox{and}\,\,  \Altb > A_{-}>A_q\,\, \hbox{if}\,\,\DDa>0,\nonumber\\
\label{ineq12}\ea
\ba \Sigma >0,\,\,\Sigltb >0\quad \hbox{if}\,\,\DDh<0,\nonumber\\
     \Sigma <0,\,\,\Sigltb <0\quad \hbox{if}\,\,\DDh>0,\nonumber\\
      \Psi_2<0,\,\,\Pltb<0\quad \hbox{if}\,\,\DDrho<0,\nonumber\\
      \Psi_2>0,\,\,\Pltb>0\quad \hbox{if}\,\,\DDrho>0.\nonumber\\
      \label{ineq34}
\ea
These inequalities, which also hold for Szekeres models compatible with PLH, lead to a qualitative but robust inference of the time preserved concavity of the radial profiles (\ref{ASigPsi3a})--(\ref{ASigPsi3b}) showing various possible sequences of maxima and minima along the radial intervals $\Delta_*^i$. This is illustrated in figure \ref{figprofiles}:
\begin{itemize}
\item Panels (a) and (b): the fluctuations $\DDa$ keep the same sign in all $\Delta_*^i$. There are $n$ saddle points of $A_q$ at the $r_*^i$, the maxima and minima of $A_+$ and $A_{-}$ lie inside the intervals $\Delta_*^i$ and at the $r_*^i$.
\item Panels (c) and (d): the fluctuations $\DDa$ alternate signs along the sequence of $\Delta_*^i$. Maxima and minima of $A_q$ occur at the $r_*^i$, while maxima and minima of $A_{+},\,A_{-}$ lie inside the intervals $\Delta_*^i$.
\end{itemize}
As the figures reveal, the order of the maxima and minima depends on the type of extrema at the origin (or the sign of $\DDa$ in the first interval $\Delta_*^1$). Evidently, any combination of the four patterns displayed can also be defined by PLH initial conditions (\ref{localhom1ab}).

PLH allows for the existence at all $t$ of an arbitrary number of extrema in the specified intervals $\Delta_*^i$ and in specified angles $[\theta_\pm,\phi_\pm]$, which follow from (\ref{sol1})--(\ref{sol2}) through a choice of dipole parameters $X,\,Y,\,Z$. In principle, even models  with an infinite number of extrema can be considered for an infinite sequence of $r_*^i$. However, simple dipole configurations (as those found in most of the literature \cite{Bsz2,NIT2011,BoCe2010,MPT,PMT,WH2012}) can also be constructed by PLH initial conditions (\ref{localhom1ab}) that admit a single Comoving Homogeneity Sphere at some $r=r_*^1$, leading to three extrema: one at the origin and the other two located in coordinates $[r_{e\pm},\theta_\pm(r_{e\pm}),\phi_\pm(r_{e\pm})]$ with $0<r_{e\pm}<r_*^1$.    

\section{Extrema from ``simulated shell crossings'' induced by the dipole parameters.}\label{noPLH}       

PLH only provides sufficient (not necessary) conditions for the existence of spatial extrema of all Szekeres scalars. As discussed in Appendix \ref{converse}, extrema for monotonic radial profiles may occur in LT and Szekeres models under special assumptions on the profiles. However, as shown in the numerical examples of \cite{Bsz2,BoSu2011,Buckley}, spatial extrema of all Szekeres scalars may occur, even without the special conditions discussed in Appendix \ref{converse}, in models defined by generic initial conditions, and thus admitting completely arbitrary radial profiles of $A_q$ and of the scalars (\ref{LTscals}) of the LT seed model. 

\begin{figure}
\begin{center}
\includegraphics[scale=0.3]{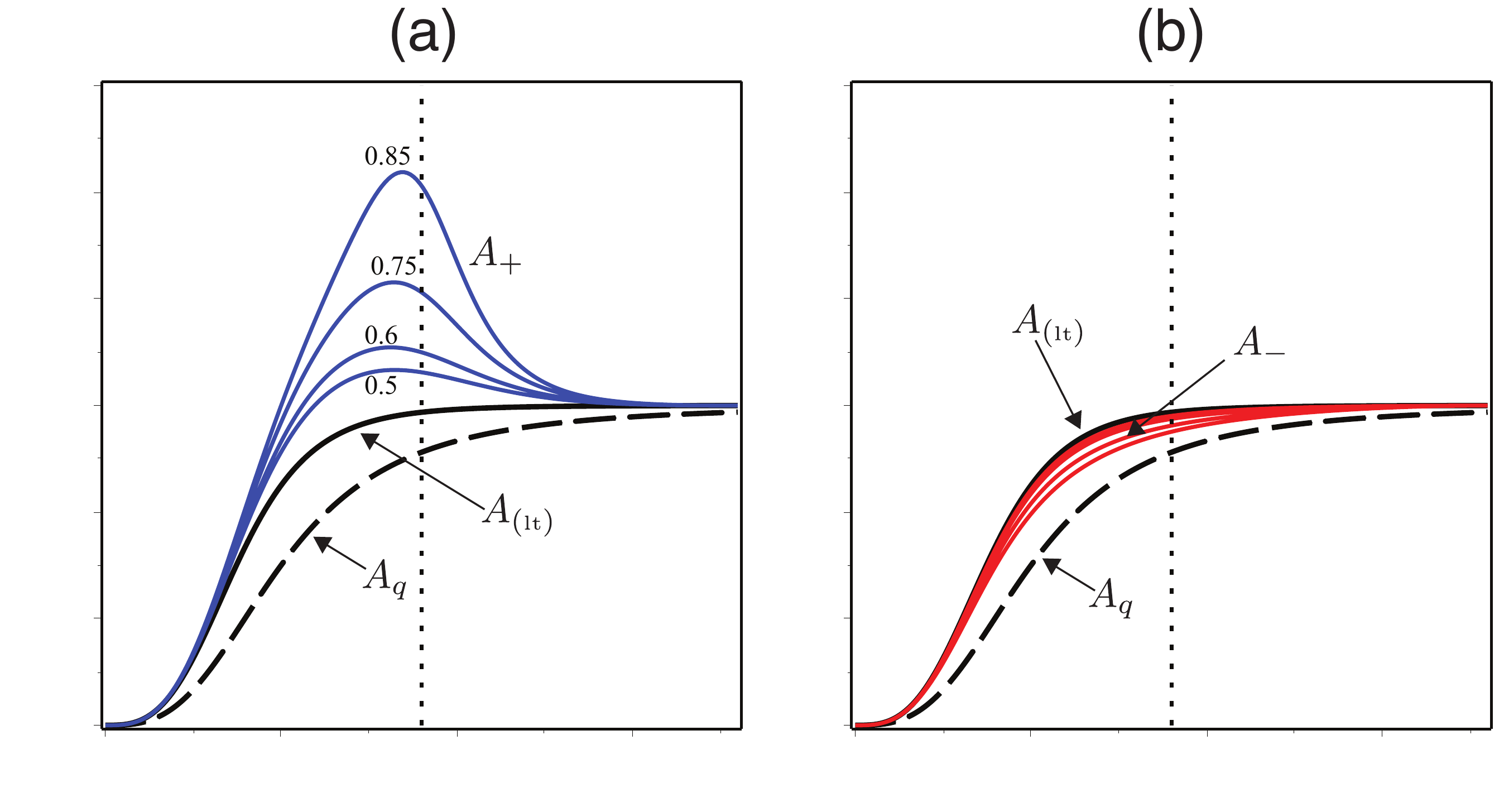}
\caption{{\bf Radial extrema from simulated shell crossings.} The panels depict the radial profiles $A_{+}$ (panel (a)) and $A_{-}$ (panel (b))  assuming a choice of dipole parameters that force a growth of $A_{+}$ for arbitrary (even monotonic) profiles of $A_q,\,\Altb$ if the term $\Gamma-\W$ takes small values close to zero (a zero marks a shell crossing singularity). We consider the time slice $t=t_0$ (hence $\Gamma_0=1$) and use the parameter choice $Z=Y=0$ (first entry of Table 1) so that $\W=X=k\sin^2[(5r/18-1)\pi]$ vanishes at $r=0$ and $r=3.6$ with $k=0.5,0.6,0.75,0.85$ controlling the deviation of $1-X$ from zero (the vertical dotted line marks $r=1.8$ where $X$ is maximal). Notice how $A_{+}$ (panel (a)) exhibits maxima with a large growth even if the radial profiles of $A_q$ and $\Altb$ are both slowly increasing monotonous void radial profiles (we used  $A_q=0.5-0.3/(1+r^3)$). }
\label{abshx}
\end{center}
\end{figure}

Generating spatial extrema in generic Szekeres models is specially straightforward for extrema along the curve $\B_{+}(r)$, since (see figure \ref{abshx}) the profiles of $A_{+},\,\Sigma_{+},\,[\Psi_2]_{+}$ grow significantly in a given radial range if the term $\Gamma-\W$ (with $\W$ given by (\ref{WW})) in the denominator of (\ref{ASigPsi3a})--(\ref{ASigPsi3b}) becomes sufficiently small for specific choices of dipole parameters $X,\,Y,\,Z$, which may occur regardless of the choice of initial value functions $A_{q0}$. Since $\Gamma-\W=0$ marks a shell crossing singularity at the curve $\B_{+}$ (see \ref{noshx}), and thus the radial profiles $A_{+},\,\Sigma_{+},\,[\Psi_2]_{+}$ diverge (which implies that the scalars themselves diverge), then a choice of dipole parameters in which $\Gamma-\W$ is positive but close to zero produces a maximum or minimum via a large growth (or decay) of these profiles (shown in figure \ref{abshx}) by creating (in specific radial ranges) conditions close to a shell crossing ({\it i.e.} regular conditions that simulate or approximate the behaviour near a shell crossing singularity).  

In order to illustrate that a maximum of $A_{+},\,\Sigma_{+},\,\Psp$ (and thus a spatial maximum along $\B_{+}(r)$) is easy to achieve from a simulated shell crossing, we consider a choice of dipole parameters such that $\Gamma-\W$ has a minimum at some $r=\rtv>0$ for a fixed $t$. Let $\Gamma(t,\rtv)-\W(\rtv)=\epsilon>0$ (for sufficiently small $\epsilon$) be the minimal value of $\Gamma-\W$, hence $\Gamma'(t,\rtv)-\W'(\rtv)=0$ necessarily holds. Under these simple generic assumptions the radial conditions (\ref{condr1}) and (\ref{condrSP}) for an extremum of either one of the radial profiles $A_{+},\,\Sigma_{+},\,\Psp$ become for $r\approx \rtv$ 
\ba  A'_{+}\approx (1+3\epsilon)A'_q(t,\rtv)+\rtv A''_q(t,\rtv)=0,\nonumber\\
\Sigma'_{+}\approx -\HH'_q(t,\rtv)-\rtv \HH''_q(t,\rtv)=0,\nonumber\\
\Psp'\approx \frac{4\pi}{3}\left[\rho'_q(t,\rtv)+\rtv \rho''_q(t,\rtv)\right]=0,\nonumber\\\label{noPLHcond1ac}\ea
and are easily satisfied  for a very wide range of generic initial conditions and radial profiles of $A_q$ in which $A'_q\ne 0$ holds but $A'_q$ and $A''_q$ have opposite signs, as this combination of signs is very common: $A_q$ grows ($A'_q>0$) with the concavity of a maximum ($A''_q<0$) or $A_q$ decays ($A'_q<0$) with the concavity of a minimum ($A''_q>0$). The type of extremum obtained by a minimum of $\Gamma-\W>0$ at $r=\rtv$ depends on the concavity of the profile of $A_q$, which is determined by the sign of $\DDa$ and follows from the type of extremum at the origin $r=0$ that is common to $A_{+},\,\Sigma_{+},\,\Psp$, but does not depend on the choice of dipole parameters because $\W(0)=0$. Hence, as illustrated in figure \ref{abshx}, if there is a central minimum of $A_{+}$ and (\ref{noPLHcond1ac}) hold, the extremum at $r=\rtv$ is necessarily a maximum (and vice versa). 

Under the conditions we have assumed,  the radial profiles of $A_{+},\,\Sigma_{+},\,\Psp$ in (\ref{ASigPsi3a})--(\ref{ASigPsi3b}) in a radial range $r\approx \rtv$ take the form
\ba A_{+} &\approx& A_q(\rtv)+\frac{\rtv\,A'_q(t,\rtv)}{3\epsilon},\nonumber\\ 
\Sigma_{+}&\approx& -\frac{\rtv\,\HH'_q(t,\rtv)}{3\epsilon},\quad 
\Psp \approx \frac{4\pi\rtv\,\rho'_q(t,\rtv)}{9\epsilon},\nonumber\\
\label{noPLHcond2}\ea
which for $\epsilon\ll 1$ clearly conveys a large growth or decay of these profiles depending on the sign of $A'_q(t,\rtv)$ (or, equivalently, the sign of $\DDa$).
%
%aborted_shx.pdf
%

An extremum of $A_{-},\,\Sigma_{-},\,\Psm$ (and thus a spatial extremum along $\B_{-}(r)$) can also occur without assuming PLH, but the conditions for this are far more stringent even if we attempt a simulated shell crossing, since following (\ref{ineq12})--(\ref{ineq34}) (which are valid even without PLH) these profiles cannot grow (or decay) more than the LT profiles (see panel (b) of figure \ref{abshx}). All this can also be appreciated from the expressions equivalent to (\ref{noPLHcond1ac}) and (\ref{noPLHcond2})
\ba  A'_{-} &\approx& \left[1+3\epsilon+6\W(\rtv)-\frac{2\rtv\W'(\rtv)}{2\W(\rtv)+\epsilon}\right]A'_q(t,\rtv)\nonumber\\
&+& \rtv A''_q(t,\rtv) = 0,\nonumber\\
 A_{-}&\approx& A_q(\rtv)+\frac{\rtv\,A'_q(t,\rtv)}{2\W(\rtv)+\epsilon}.\nonumber\\\label{noPLHcond3ab}\ea
with analogous expressions for $\Sigma_{-},\,\Psm$. Evidently, as shown in panel (b) of figure \ref{abshx}, (\ref{noPLHcond3ab}) it is much harder to satisfy than (\ref{noPLHcond1ac}) and the  growth of decay of $A_{+}$ is far more restricted than that of $A_{+}$ in (\ref{noPLHcond2}). 

Since the extrema that purely follow from simulated shell crossings are obtained by (and require) a careful choice of dipole parameters, it is very difficult (likely impossible) to provide generic sufficient conditions for their existence (as we can do by assuming PLH). However, PLH depends on the initial value functions $A_{q0}$ common to the LT seed model ({\it i.e.} conditions (\ref{localhom1ab})), whereas spatial extrema from simulated shell crossings  are completely independent of the LT initial conditions, hence generating such extrema is very useful, for example, in setting up Szekeres models constructed from specific LT seed models that comply with desired properties (as for example LT density profiles compatible with observational bounds in \cite{Bsz2,BoSu2011,Buckley}). Moreover, the practical implementation of Szekeres models admitting such extrema face the following difficulties: (i) the existence conditions (\ref{noPLHcond1ac}) change with $t$, and thus may not hold for the whole time evolution (see section \ref{inversion}); (ii) conditions (\ref{noPLHcond1ac}) must be evaluated at all $t$ to verify avoidance of real (not {\it simulated}) shell crossings. All this  depends crucially on a careful selection of the free parameters and  involves detailed numerical work. As a contrast, compatibility with PLH effectively guarantees the existence of spatial extrema for all $t$, though verifying avoidance of shell crossings still remains a technically challenging problem (see section \ref{shx}).                   
 
\section{Classificaction of the spatial extrema of Szekeres scalars.}\label{classi}

Whether we assume PLH or not, the type of the extrema of the radial profile of a scalar ({\it i.e.} the maxima and minima in figures \ref{figprofiles} and \ref{abshx}) does not (necessarily) determine the type of the  spatial extremum (maxima/minima or saddle point) of the same scalar as a full 3--dimensional object at each fixed $t$
\footnote{This fact was overlooked in references \cite{kras2,WH2012,Vrba,kokhan1,kokhan2} which mistakenly assume that maxima and minima of the radial profiles $\rho_{\pm}$ are automatically spatial maxima and minima of the density. This assumption is only correct for the extremum at the origin.}
.  To properly classify these spatial extrema (see Appendix \ref{extrA}) we need to resort to standard criteria for this purpose based on the signs of the determinants of the Hessian matrix and its minors  evaluated at the extrema location given by (\ref{extrcoords}). However, we can also obtain this classification from qualitative arguments based on the compatibility between angular maxima/minima and the maxima/minima of the radial profiles.   

\subsection{Spatial extrema at the origin.}\label{classi1}

It is straightforward to show (see (\ref{ang2derA1a}) and \ref{extrA}) that, as long as regularity conditions hold (see Appendix \ref{origin}), all second derivatives of $A$ vanish as $r\to 0$, save for the second radial derivative, which in the limit $r\to 0$ for all $t$ takes the form:
\begin{equation} \left[A''\right]_{r=0}\sim \left[A''_q\right]_{r=0}=\left[\Altb''\right]_{r=0}.\label{Arr0}\end{equation}
Since scalars have no angular dependence at $r=0$, the radial extrema are sufficient to determine the type of spatial extrema. As a consequence of (\ref{Arr0}), whether we assume PLH or not, the extrema of a Szekeres scalar $A$ at the origin $r=0$ is of the same type as the extrema of the scalar $\Altb$ of the seed LT model at the symmetry centre $r=0$:
\ba \left[\Altb''\right]_{r=0}<0\quad \hbox{local maximum of}\,\,A,\\ \left[\Altb''\right]_{r=0}>0\quad \hbox{local minimum of}\,\,A. \ea
which implies that either $\left[A\right]_{r=0}>A$ (central maximum) or $\left[A\right]_{r=0}<A$ (central minimum) hold for all points in any small neighbourhood around $r=0$. If $A''=0$ at $r=0$, it may be necessary to expand $A$ at $r\approx 0$ to higher order. 

\subsection{Spatial extrema at $r>0$.}\label{classi2}

The classification and location of the spatial 3--dimensional extrema of $A,\,\Sigma,\,\Psi_2$ at coordinates (\ref{extrcoords}) with $r_{e\pm}>0$ is as follows: 
\begin{enumerate}
\item Spatial maxima or minima only occur inside the intervals $\Delta_*^i$ along the curve $\B_{+}(r)$, as only in these locations the radial and angular maxima and minima coincide. 
\item All extrema along the curve $\B_{-}(r)$ are saddle points, as we have radial maxima vs angular minima and radial minima vs angular maxima, while extrema at $r=r_*^i$ (panels (a) and (b) of figure \ref{figprofiles}) are also saddle points because we have radial maxima/minima vs angular degenerate extrema.
\item Some spatial extrema along the curve $\B_{+}(r)$ may be saddle points.
\end{enumerate}
In what follows we prove 1 and 2 above by means of qualitative arguments based on the comparison of the concavity of the angular and radial  extrema. The case 3 is proven in Appendix \ref{extrA}.\\ 

\noindent
{\bf Classification of the angular extrema.} The concavity of the angular extrema of $A$ follows from the signs of the angular derivatives along the curves $\B_\pm(r)$
\ba  \left[A_{,\theta\theta}\right]_{e\pm} &=& \mp\frac{\W_e [\DDa]_e}{(\Gamma_e\mp\W_e)},\nonumber\\ \left[A_{,\phi\phi}\right]_{e\pm} &=& \mp\frac{(X_e^2+Y_e^2) [\DDa]_e}{(\Gamma_e\mp\W_e)\,\W_e},
\quad \left[A_{,\theta\phi}\right]_{e\pm}=0,\nonumber\\
 \label{ang2derA1a}\ea 
where  $[\,\,]_{e\pm}$ denotes evaluation at $r_{e\pm}^i$ and, to simplify the notation, $\Gamma_e$ and $\W_e$ also denote this evaluation (similar expressions arise for the angular derivatives of $\Sigma$ and $\Psi_2$). The signs of the second angular derivatives above lead to:\\

\noindent
Angular extrema of $A$ along the curve $\B_{+}$
\ba  
&{}& \hbox{minimum if}\,\,\DDa<0,\qquad \hbox{maximum if}\,\,\DDa>0,\nonumber\\
&{}& \hbox{degenerate if}\,\,\DDa=0,\label{AEtype1}\ea
Angular extrema of $A$ along the curve $\B_{-}$
\ba 
&{}& \hbox{maximum if}\,\,\DDa<0,\qquad \hbox{minimum if}\,\,\DDa>0,\nonumber\\
&{}& \hbox{degenerate if}\,\,\DDa=0,\label{AEtype2}\ea
where the usage of the term ``degenerate'' for angular extrema of Szekeres scalars is explained in section \ref{AEofA}.\\

\noindent
{\bf Classification of the radial extrema.} The concavity of the radial profiles $A_\pm$ follows from the signs of the derivatives $A''_{e\pm}$. Fortunately, under the assumptions of PLH or by generating extrema from simulated shell crossings (discussed in previous sections),  we can obtain the signs of $A''_{e\pm}$ directly from the curves displayed in figures \ref{figprofiles} and \ref{abshx} without having to evaluate this derivative.
\\

\noindent
{\bf Comparison of angular and radial extrema.} Combining the information on the signs extrema of $A''_{e\pm}$ from the curves displayed in figures \ref{figprofiles} and \ref{abshx} the information from the angular extrema summarised in (\ref{AEtype1})--(\ref{AEtype2}), we obtain for the full 3--dimensional concavity patterns in these figures:

\begin{itemize}
\item {\bf Assuming PLH: figure \ref{figprofiles}}
\begin{itemize}
\item Panel (a): In $\Delta_*^1$ we have $\DDa>0$, hence we have: (i) a radial maximum of $A_{+}$ (angular maximum) and (ii) a radial maximum of  $A_{-}$ (angular minimum). At $r=r_*^1$ we have $\DDa=0$ and two extrema: radial minima of $A_{+}$ and $A_{-}$ (both degenerate angular extrema). The pattern repeats for $\Delta_*^2,\,r_*^2,\,\Delta_*^3,\,r_*^3$ onwards.
\item Panel (b): We have an analogous pattern of extrema as in panel (a), but since $\DDa\leq 0$ the types of radial and angular extrema are switched: radial and angular minima coincide for extrema of $A_{+}$ inside the intervals $\Delta_*^i$. For all extrema at $r=r_*^i$ we have radial maxima and angular degenerate extrema. For the extrema of $A_{-}$ inside the $\Delta_*^i$ we have radial minima and angular maxima.
\item Panel (c): All extrema are inside the $\Delta_*^i$ with $\DDa$ alternating sign. In all odd numbered $\Delta_*^i$ radial maxima of $A_{+}$  coincide with angular maxima  and radial maxima of $A_{-}$ coincide with angular minima. In all even numbered $\Delta_*^i$  radial minima of $A_{+}$  coincide with angular minima and radial minima of $A_{-}$ coincide with angular maxima.
\item Panel (d): As in panel (c), but switching the behaviour of odd and even intervals $\Delta_*^i$ (since $\DDa$ has the opposite alternating pattern of signs).
\end{itemize}

\item
{\bf Simulated shell crossings without assuming PLH: figure \ref{abshx}}.

\begin{itemize}
\item Since $\DDa>0$ holds for all the radial range, there is a minimum at the origin $r=0$ and the pattern of spatial extrema is exactly as in the first interval $\Delta_*^1$ of panel (c): the radial maxima of $A_{+}$ coincide with an angular maxima and radial maxima of $A_{-}$ (if they occur) coincide with an angular minima. 
\item For the case with a maximum at the origin (not displayed) we have an analogous situation as in figure \ref{abshx}: radial and angular minima coincide only for $A_{+}$.      
\end{itemize}
\end{itemize}

\noindent
{\bf Comments.} The results 1 and 2 emerge readily from all the gathered information above. These results also follow from the rigorous classification derived in Appendix \ref{extrA}. The result 3 follows from the fact (see Appendix \ref{extrA}) that the coincidence of radial and angular maxima/minima (discussed qualitatively) is a necessary (but not sufficient) condition for a full 3--dimensional spatial maxima/minima to occur. However, pending very special forms of the free parameters, it is possible to show by looking at the behaviour of points in small neighbourhoods around the extrema that spatial extrema arise in most cases from this necessary condition, with saddle points arising when it is not satisfied (along the curve $\B_{-}(r)$ or at $r=r_*^i$).     

\section{Shell crossings and early Universe initial conditions.}\label{shx}

It is important to verify if the conditions for the existence of spatial extrema of Szekeres scalars  are  compatible with an acceptable evolution range of the models that is free from shell crossing singularities through the fulfilment of conditions (\ref{noshx1})--(\ref{noshx3}). For general initial conditions this is a technically complicated issue that must be examined numerically in a case by case basis. Hence, in order to obtain useful general conclusions we examine below avoidance of shell crossings only under the assumption of an asymptotic FLRW background and early Universe near homogeneous and near spatially flat initial conditions.   

Assuming as the initial time slice $t=t_0$ the last scattering surface,  the following constraints hold for possible FLRW backgrounds: spatially flat $\Lambda$CDM or Einstein de Sitter (EdS), or ever expanding Open FLRW models: 
\ba  \Ommi &\approx& \bar\Omega_0^m\approx 1,\,\, \Omki\approx\bar\Omega_0^k\approx 0,\nonumber\\
\rho_{q0} &\approx& \bar\rho_0,\,\, \HH_{q0}\approx\bar\HH_0,
\label{LSSa}\\
 \left|\drho_0\right| &\ll& 1,\quad \left|\dH_0\right|\ll 1,\quad \left|\Omki\dKK_0\right|\ll 1,\label{LSSb}\\
 \frac{|\drho_0|}{1-\W} &\ll& 1,\quad \frac{|\dH_0|}{1-\W}\ll 1,\quad \frac{\Omki|\dKK_0|}{1-\W}\ll 1,\label{LSSc} \ea
where $\Omki=\Ommi+\OmLi-1$ with $\Ommi,\,\OmLi$ defined in (\ref{cuadratura2}) and  $\bar\rho_0,\bar\KK_0,\bar\HH_0$ and $\bar\Omega_0^m,\bar\Omega_0^\Lambda,\bar\Omega_0^k$ are the parameters of the FLRW background at the last scattering surface, and the dimensionless relative LT fluctuations are given by
\ba \drho_0 &=& \frac{[\DDrholtb]_0}{\rho_{q0}}=\frac{r\rho'_{q0}}{3\rho_{q0}},\quad
    \dKK_0 = \frac{[\DDKKltb]_0}{\KK_{q0}}=\frac{r\KK'_{q0}}{3\KK_{q0}},\nonumber\\
     2\dH_0 &=& \Ommi\drho_0-\Omki\dKK_0.\nonumber\\
\label{relflucs}\ea   
The constraints on $\dKK_0$ depend on the FLRW background:
\ba \left|\dKK_0\right| &\ll& 1\quad \hbox{if}\,\,\KK_{q0}\to\bar\KK_0<0\,\,\hbox{or}\,\, \Omki\to \bar\Omega_0^k<1,\nonumber\\
&{}& \hbox{Open FLRW}\label{dKKas1}\\
\left|\dKK_0\right| &\sim& O(1)\quad \hbox{if}\,\,\KK_{q0}\to 0\,\,\hbox{or}\,\, \Omki\to 0,\nonumber\\ &{}&\hbox{EdS or}\,\,\Lambda\hbox{CDM}.\label{dKKas2}  \ea 
Conditions (\ref{LSSa})--(\ref{LSSc}) assure that the initial values of Szekeres scalars comply with the expected near FLRW conditions at last scattering: $\rho_0\approx\bar\rho_0,\,\HH_0\approx\bar\HH_0,\,\KK_0\approx\bar\KK_0$ and $\Sigma_0,\,[\Psi_2]_0\approx 0$. However, notice that (\ref{LSSc}) does not imply $\W\ll 1$, and thus initial Szekeres scalars need not be ``almost spherically symmetric''. 

\subsection{Ever expanding models with $\Lambda=\OmLi=0$.}\label{shx1}

Such models are characterised by $\KK_{q0}<0$ (hyperbolic) and $\KK_{q0}=0$ (parabolic). Fulfilment of (\ref{noshx1})--(\ref{noshx3}) for all $t$ implies the following constraints (which also guarantee that $\rho\geq 0$ holds everywhere) that involve only initial conditions  \cite{sussbol}
\ba 1+\drho_0 &-&\W\geq 0,\qquad 1+\frac{3}{2}\dKK_0-\W\geq 0,\label{noshxhip1}\\
\frac{r}{3} \HH_{q0} \tbb' &=& \left(\frac{3}{2}F_0-1\right)\,\dKK_0 +\left( 1- F_0\right)\,\drho_0\leq 0,\nonumber\\
&{}&\label{noshxhip2}\ea
where we are assuming that $1-\W>0$ holds with $\W$ defined in (\ref{WW}) and  $F_0=\HH_{q0}(t_0-\tbb)$ follows from (\ref{tbb}) (see analytic form in equation (66) of \cite{sussbol}). By setting $\dKK_0=0$ and $\Ommi=1$ (which implies $F_0=2/3$ from (\ref{tbb})) conditions (\ref{noshxhip1}) and (\ref{noshxhip2}) apply to parabolic models (for the latter $\tbb'<0$ must strictly hold for $r>0$).    

\subsubsection{Hyperbolic models with a simultaneous Big Bang.}\label{hypsimtbb}

Since both terms $(3/2)F_0-1$ and $1-F_0$ in (\ref{noshxhip2}) are non--negative \cite{sussmodes}, setting $\tbb'=0$ (a suppressed decaying mode \cite{sussmodes}) implies from (\ref{noshxhip2}) the following constraint (where we assumed (\ref{LSSa})--(\ref{LSSc}))
\ba  \drho_0 = -\frac{\frac{3}{2}F_0-1}{1-F_0}\dKK_0 \approx -\frac{3}{5}(1-\Ommi)\,\dKK_0,\nonumber\\
\Rightarrow\quad \hbox{sign}\left(\drho_0\right)=-\hbox{sign}\left(\dKK_0\right),\label{tbbconstr}\ea 
which considering that $\dKK_0$ and $\DDKK_0$ have opposite signs if $\KK_{q0}<0$, places extra restrictions on the PLH initial conditions (\ref{localhom1ab}). If the profile of $\rho_{q0}$ is like panel (a) of figure \ref{figprofiles}, the profile of $\KK_{q0}$ must be also like panel (a), and the same goes for profiles like panel (b). If the profile of $\rho_{q0}$ is like panel (c), then the profile of $\KK_{q0}$ must be like panel (d) and vice versa. 

If besides the general sign conditions on $\drho_0,\,\dKK_0$ that emerge from (\ref{tbbconstr}) we assume early Universe initial conditions (\ref{LSSa})--(\ref{LSSc}) and dipole parameters such that $1-\W>0$ holds everywhere, we have the following immediate conclusions: 
\begin{itemize}
\item The first condition in (\ref{noshxhip1}) is easily satisfied for any FLRW background.
\item Fulfilling the second condition in (\ref{noshxhip1}) is straightforward only for models with an asymptotic Open FLRW background (this follows from  (\ref{LSSb}) and (\ref{dKKas1}) and (\ref{tbbconstr})).
\item The fluctuation $\dKK_0$ is necessarily negative (since $\KK_{q0}\to 0$ but $\KK_{q0}<0$) and is not small (because of (\ref{dKKas2})) at least in the radial asymptotic range. Hence, for models with an asymptotic EdS background the second condition in (\ref{noshxhip1}) can be quite problematic and its fulfilment requires a careful choice of initial parameters. 
\end{itemize}
The possible shell crossings from lack of fulfilment of the second condition in (\ref{noshxhip1}) necessarily occur at asymptotic late times (see \cite{sussbol}), possibly beyond present cosmic time, and thus may be irrelevant in usual cosmological applications (examples of such shell crossings are reported for LT models in \cite{kras2,BKHC2009}).  The arguments stated above are valid whether we assume PLH or not (besides (\ref{LSSa})--(\ref{LSSb})).     

\subsubsection{Models with a non--simultaneous Big Bang.}\label{hypnonsimtbb}                             

The results described above on the first and second conditions in (\ref{noshxhip1}) remain valid if $\tbb'\ne 0$ (irrespective of whether we assume PLH or not). Hence, the basic difference is the fulfilment of condition (\ref{noshxhip2}):   
\begin{equation}\frac{r}{3} \HH_{q0} \tbb'  \approx \frac{1}{3}\drho_0+\frac{1}{5}(1-\Ommi)\,\dKK_0 \leq 0,\label{noshxhip2a}\end{equation}
which is clearly satisfied by several combinations of radial profiles that allow for positive and negative signs of the fluctuations (see \cite{RadProfs}). In particular, fulfilment of (\ref{noshxhip2a}) is favoured if $\drho_0,\,\dKK_0\leq 0$ holds, which if we assume PLH occurs for profiles of $\rho_{q0}$ and $\KK_{q0}$ given (respectively) by the forms of panels (b) and (a) of figure \ref{figprofiles}. However, for profiles with alternating signs of these fluctuations displayed in panels (c) and (d) of this figure it may be very hard to satisfy (\ref{noshxhip2a}).

\subsubsection{Early time confinement of shell crossings.}\label{confinement}  

Shell crossings that emerge from the lack of fulfilment of (\ref{noshxhip2}) ({\it i.e.} if $\tbb'>0$) necessarily occur at early times, and thus if we assume initial conditions (\ref{LSSa})--(\ref{LSSc}) these singularities can be confined to radiation dominated evolution times much earlier than the last scattering surface, where the dust source of a Szekeres model is no longer  physically acceptable. In order to illustrate this confinement of shell crossings, we assume initial conditions (\ref{LSSa})--(\ref{LSSb}) and use (\ref{Adef}), (\ref{AqDDa}), (\ref{qscals}), (\ref{DDaltb}) and (\ref{noshxhip2a}) to examine the density in the evolution range $t\approx \tbb$       
\ba \rho &=&\frac{\rho_{q0}\,(1+\drho_0-\bW)}{a^3\,(\Gamma-\bW)}\nonumber\\
&\approx& \frac{\rho_{q0}}{a^{3/2}}\,\frac{1+\drho_0-\bW}{(1+\drho_0-\bW)\,a^{3/2}-r\tbb'\HH_{q0}}\nonumber\\
 & \approx & \frac{\rho_{q0}}{a^{3/2}}\,\frac{1+\drho_0-\bW}{(1+\drho_0-\bW)\,a^{3/2}-\left[\drho_0+\frac{3}{5}(1-\Ommi)\dKK_0\right]}.\nonumber\\
 &{}&\label{rhoshx}\ea 
Since the square bracket in the denominator above is non--negative (as we assume $\tbb'>0$), a shell crossing singularity necessarily occurs as $a\to 0$ at a value
\begin{equation} a=a_{\textrm{\tiny{sx}}}=\left[\frac{\drho_0+\frac{3}{5}(1-\Ommi)\dKK_0}{1+\drho_0-\bW}\right]^{2/3}>0\label{ashxbb} \end{equation}
which depends on $(r,\theta,\phi)$ but complies with $a_{\textrm{\tiny{sx}}}\ll 1$ if conditions (\ref{LSSa})--(\ref{LSSc}) hold. Hence, the shell crossing occurs at all spatial coordinates but is entirely confined to cosmic times way before the last scattering surface where $a=a_0=1$. 

\subsubsection{Parabolic models.}

We have from (\ref{noshxhip1}) and (\ref{noshxhip2}) $-1<\drho_0\leq 0 \,\,\Rightarrow\,\,-1<\drho\leq 0$ for all $t$ \cite{sussmodes}. If we assume PLH, then extrema of $\rho$ and $\HH$ (notice that $\KK=0$) with a complete evolution without shell crossings are only possible for radial profiles $\rho_\pm$ and $\HH_\pm$ as in panel (b) of figure \ref{figprofiles}. For extrema obtained from simulated shell crossings on a monotonic radial profile, the latter cannot be as in figures \ref{abshx} and \ref{amplitudes}, but must be clump--like (decreasing with increasing $r$).  However,  if we assume early Universe initial conditions (\ref{LSSa})-(\ref{LSSb}) at $t=t_0$ then extrema from void--like profiles evolving free from shell crossings for all $t>t_0$ are possible because the shell crossings from the violation of (\ref{noshxhip2}) can be confined to times much before last scattering (section \ref{confinement}).      
      
\subsection{Ever expanding models with $\Lambda>0$.}

If $\Lambda>0$ then ever expanding models follow for initial value functions for which the cubic polynomial in the denominator of (\ref{cuadratura2}) has no positive roots, which implies $\KK_{q0}<0$ but also allows for some cases with $\KK_{q0}>0$. Since the $\Lambda$ term has negligible effect in the early time evolution, the behaviour of $\Gamma-\bW$ in the near Big Bang range $t\approx \tbb$ is the same as that discussed before for ever expanding models with $\Lambda=0$. However, the $\Lambda$ term becomes dominant in the asymptotic time range, leading from (\ref{cuadratura2}) to 
\begin{equation} \Gamma-\bW \approx 1+3\dH_0-\bW-\frac{3[(\Ommi+\OmLi)\dH_0-\frac{1}{2}\Ommi\drho_0]}{a^2},\label{GWL}\end{equation}
which leads to the following constraints given in terms of initial conditions that are  implied by (\ref{noshx1})--(\ref{noshx3}) 
\begin{equation} 1+\drho_0-\W\geq 0,\qquad 1+3\dH_0-\W\geq 0,\qquad \tbb'\leq 0.\label{noshxhipL} \end{equation}
Since the first and third condition above are identical to the first condition in (\ref{noshxhip1}) and to (\ref{noshxhip2}), all the arguments concerning these conditions for ever expanding models with $\Lambda=0$ discussed in section \ref{shx1} are valid and apply directly to ever expanding models with $\Lambda>0$. However, as opposed to the models with $\Lambda=0$, the possibility of late time shell crossings is easily avoided in models with $\Lambda>0$ by selecting early Universe initial conditions, since (\ref{LSSb})--(\ref{LSSc}) imply that the second condition in (\ref{noshxhipL}) can be satisfied for every FLRW background, including a $\Lambda$CDM background. As a consequence, if we assume initial conditions (\ref{LSSa})--(\ref{LSSc}) with a $\Lambda$CDM background, we can avoid shell crossings for all $t$ by restricting initial conditions to satisfy $\tbb'\leq 0$ in (\ref{noshxhip2a}), and if this fails we can rest assured (as in the case $\Lambda=0$) that  these singularities are confined to radiation dominated early times where the models are no longer valid (section \ref{confinement}).    

\subsection{Collapsing models and regions.} 

If $\Lambda>0$ collapsing models arise for all cases with $\KK_{q0}>0$ for which the cubic polynomial in (\ref{cuadratura2}) has a positive root, if $\Lambda=0$ then $\KK_{q0}>0$ is necessary and sufficient for a collapse. Interesting configuration emerge if the collapse occurs only in dust layers comprising a region around the origin. We examine only the case $\Lambda=0$, as analytic expressions are readily available \cite{sussbol,sussmodes} for the conditions to avoid shell crossings:
\ba  1 &+& \drho_0-\W\geq 0,\qquad \tcoll'\geq 0,\qquad \tbb'\leq 0,\nonumber\\
 &{}&\hbox{necessary and sufficient,}\label{noshxella}\\
 \drho_0 &-&\frac{3}{2}\dKK_0\geq 0,\qquad 1+3(\drho_0+\dKK_0)-\W\geq 0,\nonumber\\
 &{}& \hbox{only necessary}.\label{noshxellb} \ea
where the second condition in (\ref{noshxellb}) follows from demanding that $\Gamma-\bW>0$ holds at the ``maximal expansion'' where $\HH_q=0$ and for $\tcoll'$ we have (see equation (74) of \cite{sussbol})
\begin{equation}\frac{r}{3}\HH_{q0}\tcoll'=\frac{r}{3}\HH_{q0}\tbb' +\frac{3\pi\Ommi}{[\Ommi-1]^{3/2}}\left(\drho_0-\frac{3}{2}\dKK_0\right).\label{tcollr}\end{equation}
where the analytic form of $\tbb'$ follows from (\ref{noshxhip2}) with $F_0$ given by equation (66) of \cite{sussbol}.  Evidently, the restrictions on the compatibility between PLH initial conditions (\ref{localhom1ab}) and (\ref{noshxella})--(\ref{noshxellb}) are more stringent than in ever expanding models. Since initial conditions (\ref{localhom1ab}) imply that both $\tbb'$ and $\tcoll'$ have common zeroes with $\drho_0$ and $\dKK_0$, satisfying both sign conditions (\ref{noshxhip2}) and (\ref{tcollr}) requires a very careful choice of free parameters, more so with the alternating signs of these fluctuations in the radial profiles of panels (c) and (d) of figure \ref{figprofiles}. 

The choice of initial conditions (\ref{LSSa})--(\ref{LSSc}) eliminates shell crossings in early times (if $\tbb'=0$ or $\tbb'\leq 0$ hold) or (if $\tbb'>0$) allow for their confinement well into the radiation dominated times (section \ref{confinement}). However, these initial conditions need to be further restricted to satisfy the second constraint $\tcoll'\geq 0$ with $\tcoll'$ given by (\ref{tcollr}). In fact,  initial conditions (\ref{LSSa})--(\ref{LSSc}) require $\Ommi\approx 1$ also for positive spatial curvature, hence the gradient $r\tcoll'$ that follows from (\ref{tcollr}) can be very large even if $r\tbb'$ takes its linearised form in (\ref{noshxhip2a}). Therefore, it may be much harder to find PLH initial conditions such that shell crossings are strictly confined in times that are very close to the collapsing singularity, though collapsing models (or collapsing comoving regions) compatible with PLH may exist for which shell crossings are absent in most of the interesting ranges of their evolution. This issue requires further study and will be examined in future work.        

\section{Concavity inversions of Szekeres scalars.}\label{inversion}

We have shown that spatial extrema of all Szkeres scalars with specific concavity (maxima, minima or saddle points) arise, either through PLH initial conditions (\ref{localhom1ab}), or by selecting dipole parameters to induce simulated shell crossings.  It is important to verify if the initial concavity of these extrema is preserved, or may change, during the time evolution (we remark that such concavity inversions occur for the central extremum in generic LT models \cite{RadProfs}).  

We illustrate this concavity change taking as example the density fluctuation in the case $\Lambda=0$ (the discussion is analogous for other scalars and for the case $\Lambda>0$)
\ba  \DDrho &=& \frac{\rho_{q0}}{a^3}\frac{1+\drho_0-\Gamma}{\Gamma-\bW}\nonumber\\
 &=&\frac{\rho_{q0}}{a^3}\frac{3\left(\drho_0-\frac{3}{2}\dKK_0\right)\left[\frac{F}{F_0}-\frac{2}{3}\right]+\HH_q r\tbb'}{\Gamma-\bW}\nonumber\\
 &=& \frac{\rho_{q0}}{a^3}\frac{3\left(\drho_0-\frac{3}{2}\dKK_0\right)\frac{\HH_q}{\HH_{q0}}(F-F_0)+\drho_0}{\Gamma-\bW},\nonumber\\
 &{}&\label{DDrho}\ea
where $F$ and $F_0$ are defined in (\ref{cuadratura2})--(\ref{tbb}) and we used (\ref{noshxhip2}).  PLH initial conditions (\ref{localhom1ab}) imply $\DDrho_0=\drho_0=\dKK_0=0$ and $\DDrho=0$ for all $t$ at the values $r=r_*^i$. However, the sign of $\DDrho$ in the intervals $\Delta_*^i$ between the $r_*^i$ may change with the time evolution.  Assuming that regularity conditions (\ref{noshx1})--(\ref{noshx3}) hold (absence of shell crossings) and a non--negative $\rho_{q0}$, the conditions for this follow from the constraint $\DDrho=0$ with $\DDrho$ given by (\ref{DDrho}). We have the following cases: 
\begin{itemize}
\item If $\tbb'=0$ there are no concavity changes of $\rho$ in the intervals $\Delta_*^i$. This follows from the fact that $F/F_0-2/3$ has a definite sign for all $t$: it is positive/negative for hyperbolic/elliptic models \cite{sussmodes}. Hence, the sign of $\DDrho$ in each $\Delta_*^i$ is preserved by the time evolution, as it only depends on the sign of $\drho_0-(3/2)\dKK_0$, which is preserved for all $t$ (see details in \cite{sussmodes}).
\item If $\drho_0$ and $\dKK_0$ are both nonzero and $\tbb'\ne 0$, $\DDrho=0$ occurs if
\begin{equation} \frac{\HH_q}{\HH_{q0}}\,(F-F_0)=-\frac{\drho_0}{3\left(\drho_0-\frac{3}{2}\dKK_0\right)},\label{concinv1}\end{equation}
whose solutions (for $\drho_0\ne \frac{3}{2}\dKK_0$) are non--comoving values $\rtv=\rtv(t)$ inside the intervals $\Delta_*^i$ that may introduce concavity inversions in finite time ranges (these are the ``profile inversions with turning values'' in \cite{RadProfs}). 
\item If $\drho_0=0$ but $\dKK_0\ne 0$ (with $\tbb'\ne 0$), then $\DDrho_0=0$ but $\DDrho\ne 0$ for $t\ne t_0$: 
\begin{equation} \DDrho = -\frac{9}{2}\dKK_0\,\frac{\rho_{q0}}{a^3}\frac{\HH_q}{\HH_{q0}}\frac{F-F_0}{\Gamma-\bW},\label{DDrho3}\end{equation}
hence, for any given sign of $\dKK_0$ (assumed nonzero in the intervals $\Delta_*^i$), the sign of $\DDrho$ changes in these intervals at $t=t_0$ (where $F=F_0$): it is the opposite of the sign of $\dKK_0$ for $t<t_0$ (since $F-F_0<0$) and is the same as that of $\dKK_0$ for $t>t_0$ (since $F-F_0>0$). Hence, a concavity inversion must happen at $t=t_0$ for all $\Delta_*^i$, but the concavity at each interval is mantained for all $t>t_0$ (these are the ``profile inversions without turning values'' in \cite{RadProfs}).   
\end{itemize}
These conditions for concavity inversions of $\rho$ determine (in general) if an initial over--density (local maximum of $\rho$) can evolve into a density void (local minimum of $\rho$). Since these conditions involve only quantities of the LT seed model ($\drho_0,\,\dKK_0,\,\HH_q/\HH_{q0},\,F-F_0$),  they are the same (save for the assumption of PLH) as the conditions obtained in \cite{RadProfs} for LT models with $\Lambda=0$. The conditions for concavity inversions of the extrema need to be separately examined for the other scalars $\HH$ and $\KK$, for models that do not assume PLH and for the case $\Lambda>0$.

\section{Modelling multiple evolving cosmic structures.}\label{structures}

So far we have examined sufficient conditions for the existence and the classification of spatial extrema of all covariant scalars of Szekeres models (whether compatible with PLH or not). We now concentrate on the specific case of the spatial extrema of the density, though the results we present are directly valid and applicable to the remaining scalars. The relevance of density extrema follows from the identification:
\begin{itemize}
\item the maximal density value in an ``over--density'' occurs at a spatial maximum of $\rho$,
\item the minimal density value in a ``density void'' occurs at a spatial minimum of $\rho$,
\end{itemize}
where by ``over--density'' and ``density void'' we mean localised regions (``structures'') with significantly higher and lower density in comparison with a (suitably defined) reference or background density. Evidently, what we call ``cosmic structure'' is roughly a network of such structures at various scales and levels of coarse graining. In particular, we discuss how the spatial location of the density extrema and their amplitude (relative density growth or decay) in these structures can be specified from initial conditions.     

\subsection{Location of density spatial extrema when assuming PLH.}\label{locampl}

\subsubsection{Radial comoving location.}  

PLH constrains the radial coordinates of the spatial density extrema in each interval $\Delta_*^i$  along the curve $\B_{+}(r)$ (see figure \ref{location}) for the full time evolution (which can be free from shell crossings, see section \ref{shx}). This radial coordinate location is more precise if the values $r_*^{i-1},\,r_*^i$ that bound the intervals $\Delta_*^i$ are sufficiently close (as exemplified in figure \ref{location}). While small comoving distances $r_*^i-r_*^{i-1}$ do not (necessarily) imply (at arbitrary $t$) small physical distances in the rest frames (for example area distance $a\,(r_*^i-r_*^{i-1})$ or proper radial length), a tighter bound on the comoving coordinates of the intervals $\Delta_*^i$ is (without further assumptions) a first step to constrain the radial location of the maximal/minimal density in over--densities and voids.

\subsubsection{Angular location.}\label{angloc} 

Since PLH does not restrict the dipole parameters $X,\,Y,\,Z$, the latter become free functions to determine (through (\ref{sol1})--(\ref{sol2})) the angular location of each spatial density extremum at: $\theta=\theta_{+}(r_{e+}^i),\,\phi=\phi_{+}(r_{e+}^i)$. While suitable choices of dipole parameters $X,\,Y,\,Z$ can result in a curve $\B_{+}(r)$ that covers sufficiently spread angular directions in the space $(r,\theta,\phi)$ (see figures \ref{AEcurves} and \ref{LocHomSph}), it is more practical to define these parameters in a piecewise manner in order to prescribe a more precise angular location at each $\Delta_*^i$. 

In particular, if the Szekeres dipole $\bW$ has a single free function (see fourth entry in Table 1) then (\ref{sol1})--(\ref{sol2}) lead to  curves $\B_\pm(r)$ that are radial rays that define two constant antipodal angles: $(\theta_0,\phi_0)$ and $(\pi-\theta_0,\pi+\phi_0)$ in the coordinate space $(r,\theta,\phi)$. This ``axial--like'' dipole can be obtained if the three dipole parameters are chosen as
\begin{equation} X=-\cos\phi_0\,f,\quad Y=-\sin\phi_0\,f,\quad Z=-\cos\theta_0 f,\label{XYZone}\end{equation}
where $f=f(r)$ is an arbitrary non--negative function subjected to appropriate boundary conditions, while we have introduced the minus sign in order to identify: $(\theta_{+},\phi_{+})=(\theta_0,\phi_0)$ and $(\theta_{-},\phi_{-})=(\pi-\theta_0,\pi+\phi_0)$, so that a choice of $(\theta_0,\phi_0)$ determines the angular location of a spatial maxima or minima which are necessarily located along $\B_{+}(r)$.

\begin{figure}
\begin{center}
\includegraphics[scale=0.35]{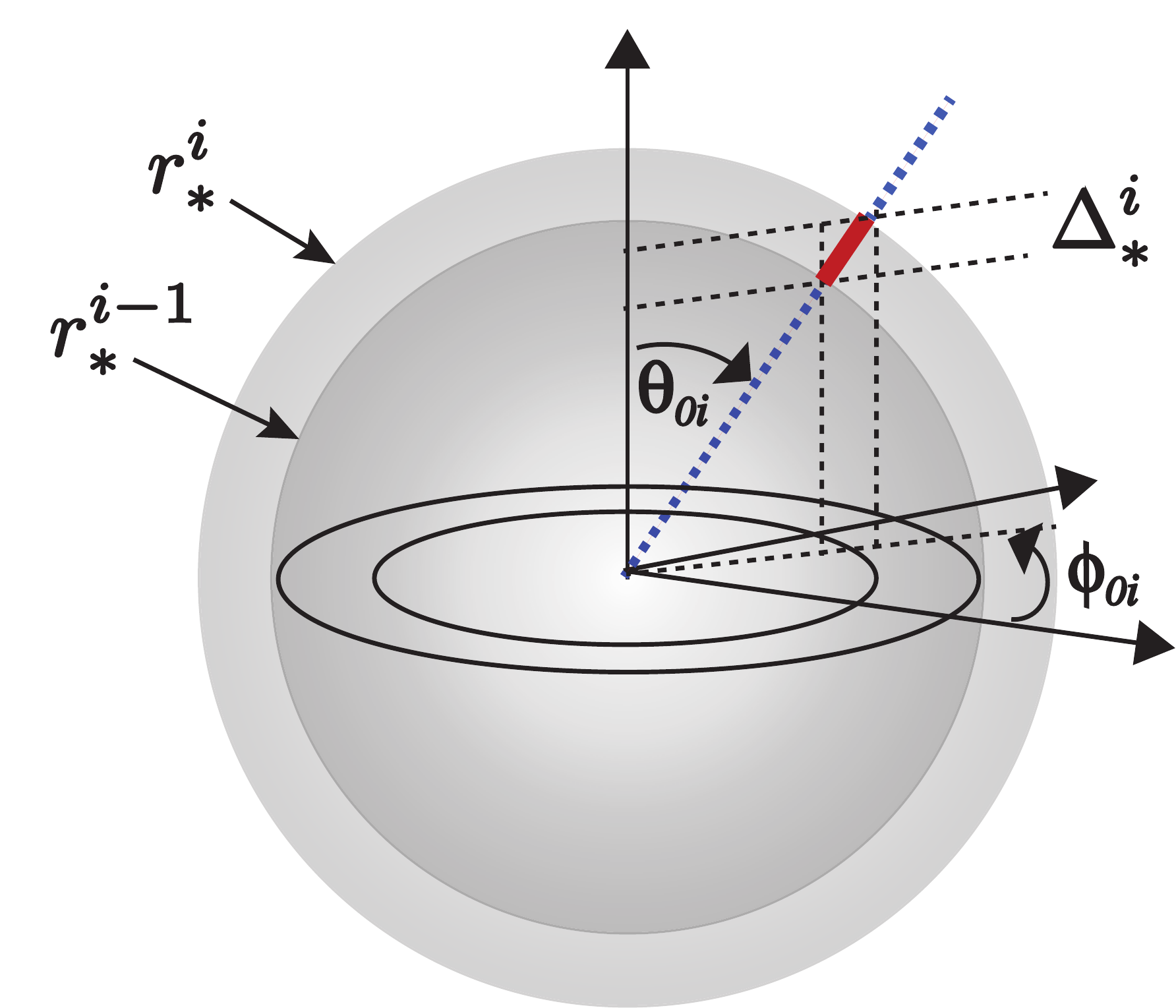}
\caption{{\bf Fixing the coordinates of the spatial maxima or minima of the density.} In Szekeres models compatible with PLH the location of these density extrema  in the coordinates $(r,\theta,\phi)$ can be fixed by: (i) selecting a radial interval $\Delta_*^i=r_*^{i-1}<r<r_*^i$, (ii) choosing the dipole parameters so  that $\B_{+}(r)$ is the thick red line segment in $\Delta_*^i$ (hence the angles $(\theta_{0i},\phi_{0i})$ specified in (\ref{pwXYZ}) are fixed in the interval, see further explanation in the text). A network of arbitrary number of spatial density maxima or minima (which identify over--densities and/or voids) can easily be set up by choosing the dipole parameters as in (\ref{pwXYZ}) so that the curve $\B_{+}(r)$ becomes a collection of piecewise continuous segments and the spatial extrema appear in desired angles in each radial interval $\Delta_*^i$ (see figure \ref{density}). This process is also possible for spatial maxima and minima of other scalars and for models not compatible with PLH involving also a suitable definition of the dipole parameters in a piecewise manner. }
\label{location}
\end{center}
\end{figure}

Considering the choice of dipole parameters explained above and given $n$ spatial maxima or minima located in $r=r_{e+}^i$ inside radial intervals $\Delta_*^i$, a convenient form to define $X,\,Y,\,Z$ is as follows
\begin{widetext}
\ba   X = 
\left\{\begin{array}{c}
 -\cos\phi_{01}\,f_1, \quad \Delta_*^1\\
-\cos\phi_{02}\,f_2, \quad  \Delta_*^2,\\
...\\
-\cos\phi_{0n}\,f_n,\quad \Delta_*^n,  
\end{array}\right.
\quad
 Y = 
\left\{\begin{array}{c}
-\sin\phi_{01}\,f_1, \quad \Delta_*^1\\
-\sin\phi_{02}\,f_2,\quad  \Delta_*^2,\\
...\\
-\sin\phi_{0n}\,f_n, \quad \Delta_*^n,  
\end{array}\right.
\quad 
Z = 
\left\{\begin{array}{c}
-\cos\theta_{01}\,f_1, \quad \Delta_*^1\\
-\cos\theta_{02}\,f_2,\quad  \Delta_*^2,\\
...\\
-\cos\theta_{0n}\,f_n, \quad \Delta_*^n,  
\end{array}\right.
\label{pwXYZ} 
\ea
\end{widetext}
where:
\begin{itemize}
\item the angles $\theta_{0i},\,\phi_{0i}$ assign to all $r$ in each radial interval $\Delta_*^i$ (which contain the extrema coordinate $r_{e+}^i$) a fixed desired angular direction (see figure \ref{location}) obtained from
\ba \theta_{+}(r)=\theta_{0i},\quad 
\phi_{+}(r)=\phi_{0i}, \qquad r\in\Delta_*^i.
\label{angles}\ea
The curve $\B_{+}(r)$ becomes a collection of piecewise continuous segments defined in each $\Delta_*^i$ (see figure \ref{density}). The curve $\B_{-}(r)$ is made of such segments located in antipodal angular directions (notice that a plus sign in (\ref{pwXYZ}) simply switches the curve segments $\B_{+}(r)$ for the curve segments $\B_{-}(r)$).   
\item the $n$ non--negative smooth functions $f_i$ must satisfy the following boundary conditions to comply with smoothness of the scalars (even for piecewise continuous curves $\B_\pm(r)$) and to prevent shell crossings:
\ba f_i(0) &=& f'_i(0)=f''_i(0)=0,\nonumber\\
    f_i(r_*^i) &=& f'_i(r_*^i)=f''_i(r_*^i)=0,\nonumber\\
     &{}& 0\leq f_i<1.\nonumber\\
     \label{bcs}\ea
Without these conditions the scalars and their first derivatives are discontinuous at the $r_*^i$ (as in the example in \cite{BoSu2011}).     
\end{itemize}
%
%location.pdf
%
%
%amplitudes_12.pdf
%
\begin{figure}
\begin{center}
\includegraphics[scale=0.32]{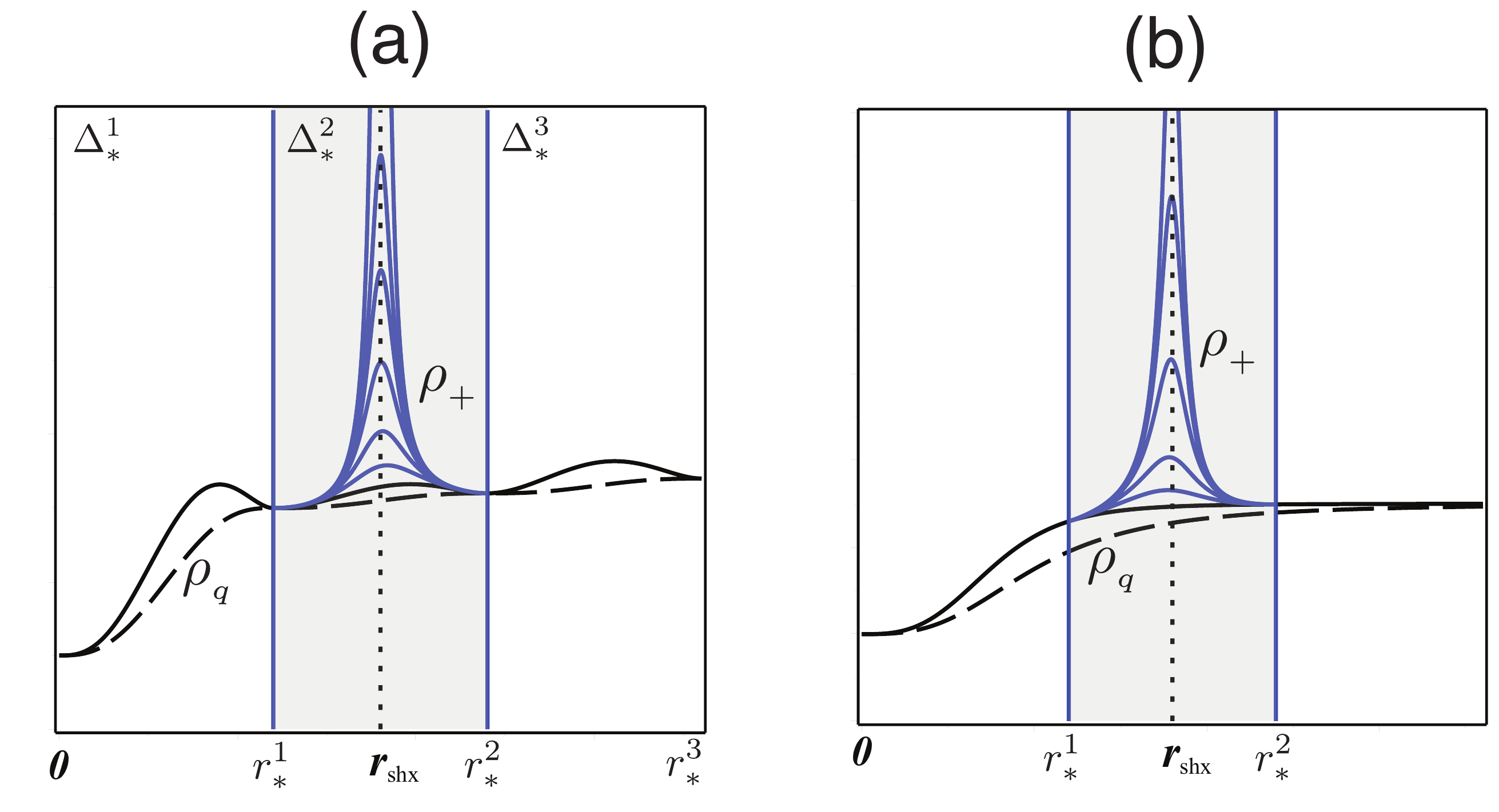}
\caption{{\bf Amplitude of a density maximum in an over--density.} The panels depict the radial profiles of $\rho_{+}$ and $\rho_q$ for the case when PLH is assumed (panel (a)) and when it is not assumed (panel (b)), for a maximum located in the interval $\Delta_*^2=r_*^1<r<r_*^2$, with $r_*^1=1.2$ and $r_*^2=2.4$. We defined the dipole parameters as explained in section \ref{amplts}, for functions $f_i$ given by (\ref{fdef}), with $k=0.3$ in the intervals $\Delta_*^1$ and $\Delta_*^3$ and $k=0.5,0.6,0.75,0.95,0.975$ in $\Delta_*^2$ (blue curves). The radial profile of $\rho_q$ in panel (a) complies with the pattern of panel (a) of figure \ref{figprofiles} and the profile in panel (b) is the same as that of figure \ref{abshx}. Notice how a larger value $k=0.975$ approaching $1$ in $\Delta_*^2$ yields very large density growth concentrated in this radial interval. The dotted vertical line is the maximum of $f_2$ in $\Delta_*^2$, which almost coincides with the radial maximum of $\rho_{+}$. If $k=1.0$, this value would mark a shell crossing singularity in which $\rho_{+}$ diverges.}
\label{amplitudes}
\end{center}
\end{figure}
\subsection{Amplitude.}\label{amplts} 

While it is possible to control the density growth or decay at each over--density or density void in an interval $\Delta_*^i$ from the choice of initial conditions (\ref{localhom1ab}), it is far easier to control this growth or decay by the simulated shell crossing effect through a suitable choice of dipole parameters, as discussed in section \ref{noPLH} and illustrated by figure \ref{abshx}. Consider a single interval $\Delta_*^i$ in a radial profile $\rho_{+0}$ at the initial time $t=t_0$ given by (\ref{ASigPsi3a}) and following the pattern displayed in panel (a) of figure \ref{figprofiles}, which yields a spatial minimum at $r=0$ and a sequence of $n$ spatial maxima at each $\Delta_*^i$. For simplicity we assume dipole parameters as in (\ref{pwXYZ}) with $\phi_{0i}=0,\,\theta_{0i}=\pi/2$ (first entry of Table 1), so that $Y=Z=0$ and $X=-f_i$ hold in $\Delta_*^i$. A simple convenient form for $f_i$ that satisfies the boundary conditions (\ref{bcs}) is 
\begin{equation} f_i=k_i\,\sin^2\left[\frac{(r-r_*^{i-1})\pi}{r_*^i-r_*^{i-1}}\right],\qquad \W=|X|=\,f_i,\label{fdef}\end{equation}
where $0<k_i<1$, which allows us to use $k_i$ to control the growth of the density profile $\rho_{+}$ in $\Delta_*^i$ as depicted in panel (a) of figure \ref{amplitudes}. Comparison of this figure with figure \ref{abshx} reveals that a more ``narrow'' radial range produces a more localised simulated shell crossing effect, which allows us to control the density growth of decay more effectively. Evidently, this process of setting up the amplitude of the over--density in figure \ref{amplitudes} can be applied for the density extrema in the remaining over--densities at every other $\Delta_*^i$ (and to other Szekeres scalars). In fact, it can be applied to every combination of over--densities and voids in the $\Delta_*^i$ that follows from the profiles of  displayed in figure \ref{figprofiles} and arbitrary combinations of them.     

\subsection{Models not compatible with PLH.}

As discussed in section \ref{noPLH}, spatial extrema of the density that identify over--densities or voids can be generated by the simulated shell crossing effect on arbitrary radial profiles, leading (for monotonic profiles of the LT density) to the two pattern of maxima and minima described in section \ref{classi}. In particular, any regular Szekeres model admits dipole parameters defined as in (\ref{pwXYZ}), hence we also set up networks of over--densities or voids in assorted radial and angular locations and can control their amplitude by choices of functions $f_i$ complying  with (\ref{bcs}). Panel (b) of figure \ref{amplitudes} illustrates setting up a density maximum at a given interval $\Delta_*^i$ when PLH is not assumed. 

\section{Numerical Example.}\label{numex}

Szekeres models contain sufficient degrees of freedom to allow for setting up very elaborated and complex networks of evolving over--densities and density voids. If we assume PLH, the general morphology of these networks is also preserved (pending possible shell crossings or concavity inversions discussed in sections \ref{shx} and \ref{inversion}) by the time evolution, as PLH is preserved by the time evolution (see numerical examples of these evolving structures in \cite{nuevo}).
%
%density_ai.pdf 
%
\begin{figure*}
\begin{center}
\includegraphics[scale=0.40]{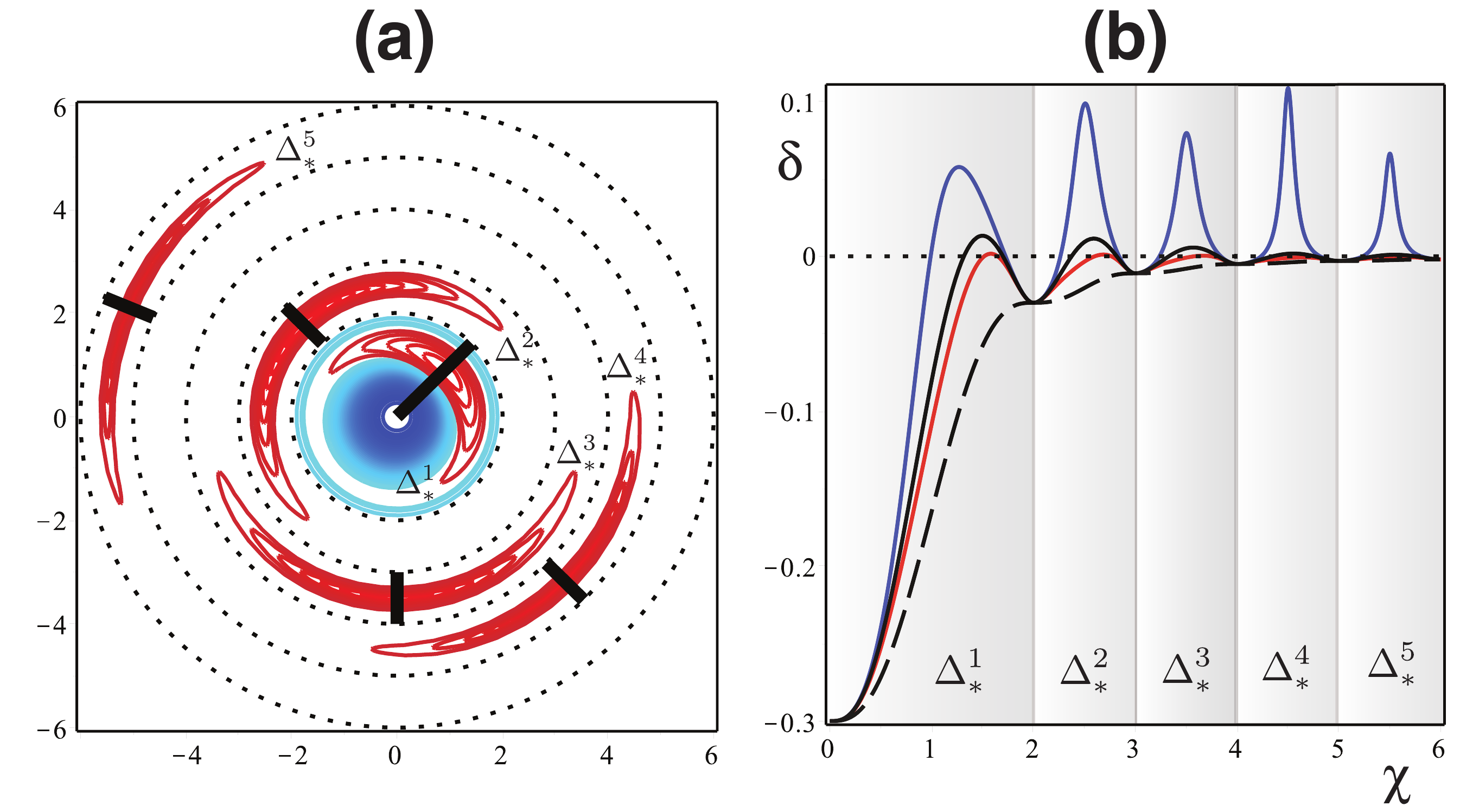}
\caption{{\bf Void with five over--densities.} Panel (a) depicts the level curves (in units of $10^{-4}$) in the equatorial plane ($\theta=\pi/2$) of an initial density contrast $\delta=\rho_0/\bar\rho_0-1$ at the last scattering surface as a function of $\chi=r/r_s$ and $\phi$, where $\bar\rho_0$ is the background density and $r_s$ is a convenient subhorizon scale. We assumed a profile of $\rho_{q0}$ as in panel (a) of figure \ref{figprofiles} for dipole parameters chosen as piecewise functions as in (\ref{pwXYZ}) with $\theta_{0i}=\pi/2$ (so that $Z=0$) and all $f_i$ given by (\ref{fdef}) with $k=0.75,0.8,0.85,0.9,0.9$. The spheroidal density void around the origin is depicted as a blue--cyan area corresponding to contrast values $-0.3<\delta<-0.01$ (colours only appear in the online version). The void is surrounded by four sharply defined over--densities (orange red areas corresponding to $0.01<\delta<1.1$), each one located around a density maximum in an interval $\Delta_*^i$ in the angular locations $\phi_{0i}=\pi/4,3\pi/4,3\pi/2,7\pi/4,7\pi/8$. The circles correspond to the Comoving Homogeneity Spheres marked by coordinates $r_*^i$, while the thick black segments are the curves $\B_{+}(r)$. The over--densities are filamentary shaped in both angular directions $(\theta,\phi)$, hence their 3--dimensional morphology is pancake like (see figure \ref{pancake}). Panel (b) depicts the radial profile of $\delta$ along $\B_{+}(r)$ (blue curve) and $\B_{-}(r)$ (red curve) together with the profiles of $\rho_{q0}$ (black dashed curve), the LT density contrast (solid black curve) and the zero contrast background value (dotted horizontal line). This type of configurations are preserved by the time evolution (see numerical examples in \cite{nuevo}).}
\label{density}
\end{center}
\end{figure*}
%
%over_density_joint.pdf
%
\begin{figure}
\begin{center}
\includegraphics[scale=0.38]{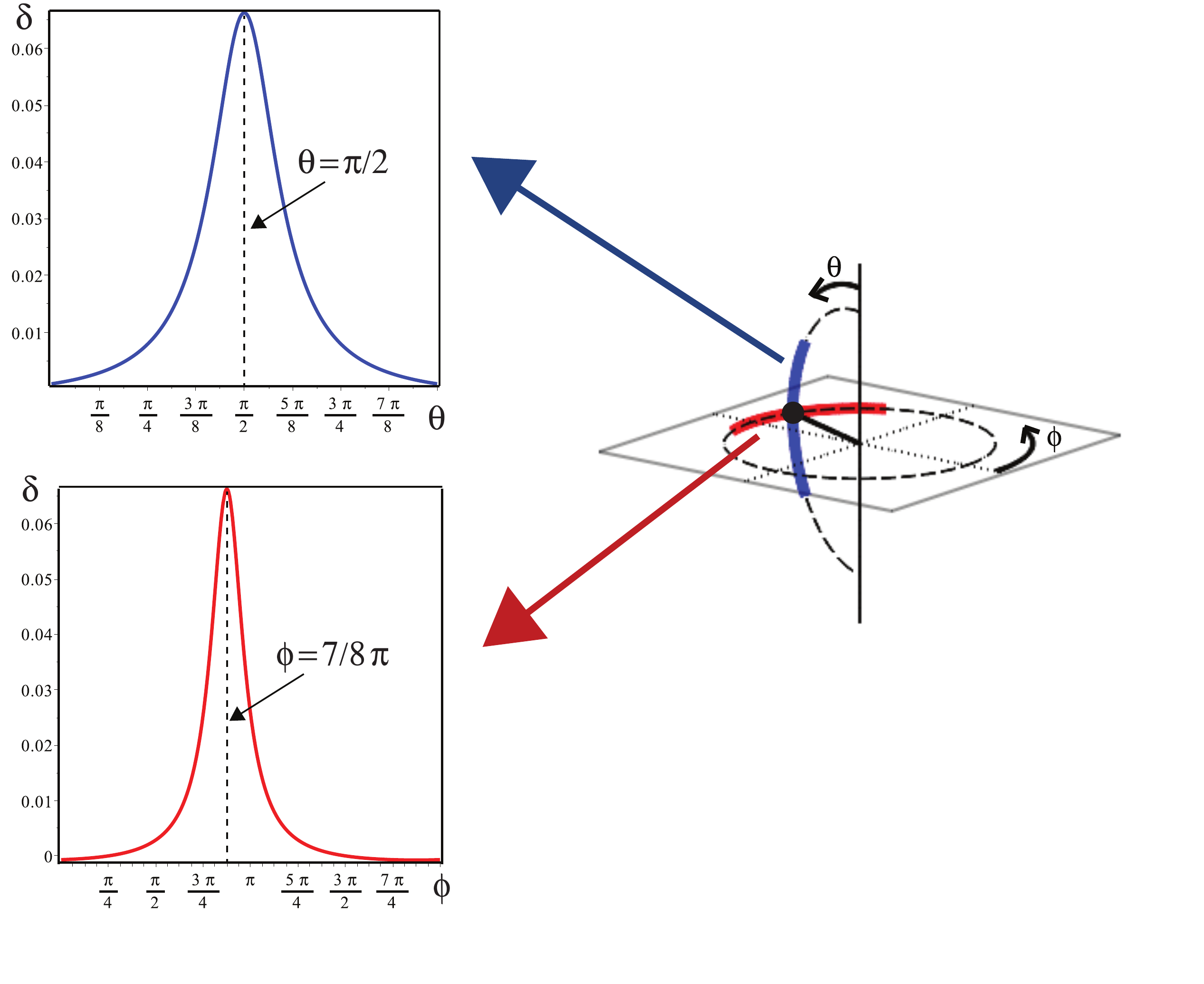}
\caption{{\bf Thick pancake morphology of over--densities.} The figure illustrates the 3--dimensional ``thick pancake'' shape of the over--densities (with extrema not located in $r=0$) that arise from Szekeres models. We use as example the over--density of figure \ref{density} around the density maximum located in the interval $\Delta_*^5$ in angular coordinates $\theta_{05}=\pi/2,\,\phi_{05}=7\pi/8$. The right hand side diagram shows how this over--density spreads along both angular directions $\phi$ (red) and $\theta$ (blue) in the coordinate space $(r,\theta,\phi)$. The plots on the left side show the density contrast as a function of $\theta$ (for $\phi_{05}=7\pi/8$) and $\phi$ (for $\theta_{05}=\pi/2$), with the maximal values at the angular coordinates of the density spatial maximum. The radial density variation is displayed in panel (b) of figure \ref{density}. From the angular and radial spread of the density around the spatial maximum we can conclude that the over--density has roughly a thick pancake morphology.  }
\label{pancake}
\end{center}
\end{figure}

\subsection{Networks of structures.}

We provide in this section a simple numerical example of a network of structures that is displayed in figure \ref{density}. Panel (a) of this figure depicts the equatorial projection of the level curves of the density contrast $\delta=\rho_0/\bar\rho_0-1$ (with $\bar\rho_0$ the background density) of an initial density  configuration at the last scattering surface (hence $|\delta|<10^{-3}$). This configuration consists of a large spheroidal shaped central void surrounded by five sharply defined elongated over--densities obtained from PLH initial conditions (\ref{localhom1ab}) with the profile of $\rho_{q0}$ as in panel (a) of figure \ref{figprofiles}. We selected the dipole parameters as $Z=0$ ($\theta_{0i}=\pi/2$ for all $i$, fifth entry of Table 1), with $X,\,Y$ given as in (\ref{pwXYZ}) to determine a unique angular location for the spatial density maximum at each over--density in the equatorial plane. As shown in panel (b) of figure \ref{density}, the amplitude of the density contrast for each over--density is controlled by the radial profile of $\rho_{q0}$ and by the functional form of $X$ and $Y$ (see section \ref{noPLH}).

\subsection{Morphology and evolution of the structures.} 

The shape of the central void in panel (a) of figure \ref{density} indicates a spheroidal morphology, which (from qualitative arguments) should hold for every structure (over--density or void) around a density extremum at the origin. As a contrast, the over--densities in panel (a) of figure \ref{density} look like elongated thin filaments, a shape that follows from the  fact that this panel only displays the radial and azimuthal ($\phi$ direction) density spread in the equatorial projection ($\theta=\pi/2$) of the density contrast, and thus it is insufficient to convey the the full 3--dimensional morphology of these over--densities. The missing information is provided in figure \ref{pancake}, using as example the over--density located in the interval $\Delta_*^5$ of figure \ref{density} (the same results hold qualitatively for all over--densities not located at $r=0$). The right hand side diagram of figure \ref{pancake} shows (qualitatively) how the density spreads around the density maximum along both, the ``meridians'' (the $\theta$ direction) and the ``parallels'' (the $\phi$ direction) of the spherical coordinate system. The left hand side plots of this figure show the angular density profiles (along both angular directions) at the fixed radial coordinate of the density maximum. Combining the information of these graphs and the graph of the radial profile in panel (b) of figure \ref{density}, we can conclude that the general 3--dimensional morphology of these over--densities is roughly that of ``thick pancakes'', whose ``thickness'' depends on the ratio of physical distances along the different spatial directions in which the density is larger that an given background reference value.     
 
While we obtained the numerical example of figures \ref{density} and \ref{pancake} by assuming PLH, qualitatively analogous structures can also be obtained without assuming this property, by generating maxima (or minima) of the density through simulated shell crossings and then selecting dipole parameters as explained in section \ref{structures}. This procedure to generate a large number of over--densities in the equatorial plane was used in \cite{BoSu2011}, but the dipole parameters $X$ and $Y$ (defined as in (\ref{pwXYZ})) were not continuous in the boundaries separating each piecewise radial range, which leads to discontinuous density (a ``thin shell'' approximation). As we show in the example depicted in figure \ref{density}, this inconvenience can easily be fixed by a suitable choice of dipole parameters satisfying appropriate boundary conditions (see section \ref{location}).

As a general rule that emerges from figures \ref{density} and \ref{pancake}, the most interesting type of cosmic structure generated by Szeleres models (assuming PLH or not) consists of a density minimum at $r=0$ surrounded by a central spheroidal underdense region (density void), which is in turn surrounded by an array of curved thick pancake shaped over--densities whose density maxima can be located in assorted radial and angular directions that can be prescribed by initial conditions. Evidently, by using the full degrees of parameter freedom described in section \ref{angloc}, networks of pancake--like over--densities around spheroidal voids can be generated that are more general than those of figure \ref{density}, with the density maxima in the over--densities located all over the 3--dimensional space. Also, if we need to construct a mixed network of pancake--like over--densities and voids, then we can  choose for $\rho_{q0}$ and $\KK_{q0}$ following different combinations of the radial profiles displayed by figure \ref{figprofiles} that comply with absence of shell crossings at least for the evolution range $t>t_0$ (see section \ref{shx}).      

Figure \ref{LocHomSph} provides a rough qualitative description of the spacetime evolution of initial density configuration like the one displayed in figure \ref{density}, showing how (under the assumption of PLH) each over--density evolves in the spacetime shell regions between Comoving Homogeneity Spheres. As we showed in section \ref{shx}, this evolution can be free from shell crossings for a wide variety of early Universe initial conditions as those used in figure \ref{density}. However, without assuming PLH it is not evident if a given initial concavity (spatial maxima or minima) is preserved for the whole evolution, or if concavity inversions may arise (see section \ref{inversion}). All this must be examined through a proper study of the time evolution of these configurations, which requires a separate article (we address these issues in \cite{nuevo} and will address them also in future work).       

\section{Summary and final discussion}\label{final}

We have undertaken a comprehensive unifying study of the existence, classification and location of spatial extrema (local maxima, minima and saddle points at hypersurfaces of constant $t$) of the main covariant scalars of Szekeres models: $A=\rho,\,\HH,\,\KK$ (density, Hubble expansion and spatial curvature) and the eigenvalues $\Sigma,\,\Psi_2$ of the shear and electric Weyl tensor.
We have expressed these scalars in terms of covariant variables \cite{sussbol} (the q--scalars and their fluctuations section \ref{sphercoords}) that reduce in their linear limit to standard variables of cosmological perturbations \cite{perts}. 

\subsection{Radial and angular extrema.}

We have used spatial spherical coordinates that lead to a complicated non--diagonal metric, but are very useful for studying the spatial location and the existence conditions for spatial extrema of all Szekeres scalars, as we can split these conditions (section \ref{locextr}) into interdependent separate conditions for the angular and radial extrema and then look at the full 3--dimensional picture. This process lead to the following results: 
\begin{itemize}
\item The Szekeres dipole $\bW$ defines at each 2--sphere of constant $r$ a precise angular coordinate location $\theta=\theta_\pm(r),\,\phi=\phi_\pm(r)$ through its {\bf angular extrema} that follow from the condition $\bW_{\theta}=\bW_{\phi}=0$. For all $r$ these angular extrema define two curves $\B_\pm(r)=[r,\theta_\pm,\phi_\pm]$ in the coordinate space $(r,\theta,\phi)$ whose functional form at all $t$ is determined by the three dipole parameters $X,\,Y,\,Z$ (see section \ref{Szdip}).     
\item The angular extrema of all Szekeres scalars coincide with the angular extrema of the Szekeres dipole (see section \ref{AEofA} and figures \ref{AEcurves} and \ref{LocHomSph}). Therefore, finding the location of the full 3--dimensional extrema (the ``spatial'' extrema) reduces to solving the radial extrema conditions $A'_\pm=0,\,\Sigma'_\pm=0,\,\Pspm'=0$, where $A_\pm,\,\Sigma_\pm,\,\Pspm$ are the ``radial profiles'' obtained by evaluating the Szekeres scalars $A,\,\Sigma,\,\Psi_2$ along the curves $\B_\pm(r)$ (see section \ref{perfiles} and figure \ref{figprofiles}). In general, the radial location of the spatial extrema along these curves shifts in time ({\it i.e.} the worldlines of spatial extrema are not comoving).
\item The origin $r=0$ (assuming that regularity conditions hold, see Appendix \ref{origin}) marks for all times the location of a spatial extremum, which (see section \ref{classi1}) has the same concavity (local maximum or minimum) of the extremum of the LT seed model at its symmetry centre.    
\end{itemize}

\subsection{Existence conditions and classification of spatial extrema.}

Without solving the radial extrema equations (which requires numerical work), we have looked at sufficient conditions for the existence of spatial extrema of all scalars (section \ref{PLH} and proof in Appendix \ref{formalproofs}). In particular, we have proven that arbitrary numbers of spatial extrema for each Szekeres scalar arise by assuming compatibility of the models with {\bf{Periodic Local Homogeneity}} (PLH), defined by  demanding that the shear and Weyl tensors vanish along a sequence of $n$ evolving 2--spheres (Comoving Homogeneity Spheres, see figure \ref{LocHomSph}). PLH is preserved by the time evolution (pending absence of shell crossings) and can be specified by suitable (PLH) initial conditions that define specific patterns of radial profiles (figure \ref{figprofiles}) which are helpful to understand qualitatively the conditions for the existence and classification of the spatial extrema.    

We have also examined how spatial extrema of all Szekeres scalars can arise without assuming PLH. One  possibility (discussed in section \ref{noPLH} and illustrated in figure \ref{abshx}) follows by setting up the dipole parameters to induce a large growth (or decay) of the scalars in specific radial and angular coordinate ranges. We call this procedure ``generating spatial extrema  by {\bf{simulated shell crossings}}'', as it involves approximating the behaviour of the scalars near a shell crossing singularity, but carefully setting up the free parameters so that the actual singularity is avoided.           

A rigorous classification of the spatial extrema (whether PLH is assumed or not) is given in Appendix \ref{extrA}. However, this classification can also be obtained qualitatively (section \ref{classi}) by assuming that an extremum is a spatial maximum or minimum if its radial and angular concavities coincide, while spatial saddles occur whenever these concavities do not coincide. For Szekeres models compatible with PLH (section \ref{PLH}), the concavity of the $2n+1$ spatial extrema of every Szekeres scalar is given as follows:
\begin{itemize}
\item spatial maxima and minima only occur in the origin and in each one of the radial intervals $\Delta_*^i$ (between Comoving Homogeneity Spheres) along the curve $\B_{+}(r)$ in the following combinations (see details in figure \ref{figprofiles} and in the text of section \ref{PLH}):
\begin{itemize}
\item a spatial minimum at the origin and an array of $n$ spatial maxima.
\item a spatial maximum at the origin and an array of $n$ spatial minima..
\item a spatial maxima or minima at the origin and any type of mixed array of $n$ spatial maxima and minima.   
\end{itemize}
\item along the curve $\B_{-}(r)$ we have necessarily $n$ spatial saddle points,  
\item all extrema located at the Comoving Homogeneity Spheres are spatial saddle points.   
\end{itemize} 
For Szekeres models not complying with PLH, whose extrema may follow from simulated shell crossings (see section \ref{noPLH}), we have the following combinations that are independent of  the profiles of the scalars of the LT seed model
\begin{itemize}
\item spatial minimum in the origin $r=0$ and possible arrays of spatial maxima at $r>0$ (see figure \ref{abshx})
\item spatial maximum in the origin $r=0$ and possible arrays of spatial minima at $r>0$.
\end{itemize} 
These are general conclusions that may admit exceptions, since (as we show in Appendix \ref{extrA}) the coincidence of radial and angular concavities is a necessary but not sufficient condition to determine the concavity of the full 3--dimensional extrema.

\subsection{Shell crossings and concavity inversions. }

We examined in section \ref{shx} the compatibility between the conditions to avoid shell crossings (\ref{noshx}) and the conditions for the existence of spatial extrema. By assuming an asymptotic FLRW background and selecting standard (near homogeneous near spatially flat) initial conditions at the last scattering surface $t=t_0$, we found the following results for models admitting spatial extrema (assuming PLH or not):
\begin{itemize}
\item For ever expanding models (with $\Lambda=0$ and $\Lambda>0$) an evolution free from shell crossings for all times is possible, with some restrictions on the initial conditions for PLH. The exception is hyperbolic models with an Einstein de Sitter background in which late time shell crossings may occur regardless of the assumption of PLH. 
\item Some PLH initial conditions may induce early times shell crossings, either with a simultaneous Big Bang $\tbb'=0$ or when the condition $\tbb'\leq 0$ is violated. These shell crossings can be confined to evolution times much before the last scattering, where the models are no longer valid (section \ref{confinement}). 
\item Avoidance of shell crossings is much harder to achieve for collapsing (elliptic) models or regions that admit spatial extrema of Szkeres scalars (whether PLH is assumed or not). Shell crossings may be eliminated or confined to times before last scattering, but it is very difficult at the same time to avoid shell crossings in the late collapsing stage or near the collapsing singularity. However, reasonable long evolution times without shell crossings may be possible.  
\end{itemize}
Another important technical issue  is the possibility of concavity ``inversions'' of the spatial extrema ({\it i.e.} a local maximum evolving into a local minimum, section \ref{inversion}). Some of these inversions occur during a limited time range and only for a single scalar (the density but not the Hubble expansion or vice versa). Other possible inversions occur at a given time slice (see comprehensive study for LT models in \cite{RadProfs}).

\subsection{Modelling cosmic structure.} 

We examined the specific case of the spatial maxima and minima of the density (section \ref{structures}), which can be associated with ``structures'' such as over--densities or density voids. Whether we assume compatibility with PLH or not, Szekeres models allow for the existence of elaborated networks of arbitrary numbers of such structures around density maxima or minima whose spatial location can be  specified by initial conditions (section \ref{locampl}) and whose ``amplitude'' (relative density growth) can be controlled by the choice of dipole parameters following the simulated shell crossings procedure  (section \ref{amplts}). 

We have presented a simple numerical graphic example (displayed in figures \ref{density} and \ref{pancake}) of a configuration consisting of a central spheroidal void surrounded by over--densities whose full 3--dimensional morphology is that of thick curved pancakes (the combination of central over--density and pancake voids or any other mixed arrangements are also possible). Since we can prescribe the comoving spatial location and amplitudes of the density maxima at each over--density in such networks, we do obtain a sophisticated toy model that has the potential to realise a coarse grained fitting of observed large scale cosmic structure \cite{cosmography1,cosmography2,struct,superstruct}. While we have shown that such networks are preserved in time under the assumption of PLH, the details of their evolution requires separate articles (see \cite{nuevo}). 

It is important to remark that all conclusions we have found on the spatial maxima and minima (and saddles) of the density apply to the spatial extrema of the remaining covariant scalars, which may be useful for a full theoretical understanding of the models and their cosmological applications.            

\subsection{Improvements on previous literature.}         

There are various elements that intersect the contents of our work among the literature on Szekeres models, specially in references \cite{kras2,BKHC2009,Bsz1,Bsz2,PMT,BoSu2011,sussbol,WH2012,Buckley,Vrba,kokhan1,kokhan2}. However, in several key issues our results  represent a significant improvement over this existing work.  Firstly, all these references (including for example \cite{kras2,BKHC2009,Bsz1,Bsz2,WH2012,Vrba,kokhan1,kokhan2}) have only looked at the maxima and minima of the density (over--densities and density voids), whereas we have considered a broad theoretically unifying study of generic spatial extrema (including saddles) of all the covariant scalars of the models. We regard this as a specially relevant improvement on previous work, as the extrema of the Hubble scalar $\HH$ or the shear eigenvalue $\Sigma$ may be related in a non--trivial manner to peculiar velocities and the Hubble flow of these structures. Secondly, we have produced a clear systematic self--consistent procedure based on rigorous sufficient existence conditions and classification of an arbitrary number of extrema of all Szekeres scalars, as well as on the ability to prescribe their spatial location (at all times) of  from initial conditions that prevent shell crossings at least for the evolution range where the models are valid.  

It is important to comment on references \cite{kras2,BKHC2009,WH2012,Vrba}, which have also looked at the existence of density extrema and their compatibility with absence of shell crossings. The approach of these references to the density extrema is very rudimentary and incomplete: they have (specially \cite{Vrba}) only examined the extrema of the radial profiles $\rho_\pm$ (see definition in section \ref{perfiles}) and have thus assumed (without proof) that these are spatial maxima or minima of the full 3--dimensional density (the same remarks apply to the intention of identifying structures from the ``onion model'' and the ``mock n--body simulation data'' of \cite{kokhan1,kokhan2}). However, as we showed in section \ref{classi} (see rigorous proof in Appendix \ref{extrA}) this is only correct for the extrema at the origin, but is completely mistaken for the remaining extrema: only the maxima or minima of the radial profile $\rho_{+}$ (at $r>0$ along $\B_{+}(r)$) are spatial maxima and minima, while all the extrema of $\rho_{-}$ (at $r>0$ and along $\B_{-}(r)$) are saddles. Here we need to point out that the standard coordinates and metric functions  used by these references (see Appendix \ref{usual}) makes it much more difficult to obtain a proper classification of the spatial extrema of all scalars (including the density).   

Various aspects of our results have been discussed in previous articles following a different methodology. References \cite{WH2012,Buckley} also introduced the notion of an LT seed model and the dipole parameters are also defined as piecewise functions in \cite{Bsz2,BoSu2011,Buckley}, but all this has been carried on without our systematic and extensive approach. Theoretical and observational issues were examined in  \cite{Bsz1,Bsz2,KrBo2010,PMT,BoSu2011,sussbol} on models that describe, either an over--density next to a void \cite{Bsz1,Bsz2,PMT,Buckley} or  multiple structures \cite{BoSu2011,Buckley}, but the procedure to determine their spatial location is not systematic and is not explained with sufficient methodological rigour. While a nice numerical example of multiple over--densities was produced in \cite{BoSu2011} based on \cite{Bsz1,Bsz2}, the piecewise defined dipole parameters were discontinuous, which lead to a discontinuous density (as a contrast, we have shown how to define these parameters in order to guarantee that all scalars are everywhere smooth). The results we have obtained in this article are based on a more solid theoretical methodology, and thus greatly enhance and complement all this previous literature.

\subsection{Potential applications and limitations.}   

In spite of their limitations (which we discuss further ahead), the Szekeres configurations that we have presented provide a fully relativistic non--perturbative and non--linear description of (still) idealised but non--trivial evolving CDM structures. The following is a (by no means exhaustive) list of potential theoretical and empirical applications for these configurations:
\begin{itemize}
\item {\bf Modelling cosmic structure and fitting observations.} The multi--structure configurations can be applied to obtain a coarse modelling of existing large scale structure (from galactic surveys \cite{cosmography1,cosmography2,struct,superstruct} or numerical simulations \cite{simu1,simu2}), thus improving previous work based on spherical LT models (whether local structure \cite{BKHC2009,BolHel} or Gpc sized voids \cite{BKHC2009,BCK2011,marranot,bisnotval}) or on simple Szekeres models describing an over--density next a density void in a crude dipolar array or `Swiss cheese'' arrays \cite{Bsz1,Bsz2,NIT2011,BoCe2010,MPT,PMT,WH2012}. The networks of over--densities and voids that follow from the class I models we have examined are also an improvement over the plane symmetric ``pancake'' like structures that emerge from Szekeres models of class II \cite{kasai,meures} (the difference between class I and II models is discussed in \cite{kras2}). Our work can also provide theoretical support to the observational studies of \cite{KrBo2010,kokhan1,kokhan2}.   
\item {\bf Relativistic corrections.} By being fully relativistic and non--perturbative, the multi--structure Szekeres configurations are ideally suited to examine the effects of relativistic and non--linear corrections to Newtonian or perturbative structure modelling or observations fitting \cite{Grcorr1,GRcorr2}. In particular, our results can be relevant to complement recent work in \cite{kokhan1,kokhan2}.
\item {\bf Theoretical issues.} By being obtained from an exact non--spherical solution of Einstein's equations, the multiple structure models are ideal theoretical tools to test ideas and concepts that involve non--perturbative spacetime inhomogeneity, such as cosmic censorship and apparent horizons \cite{Nolan,krasbolAH}, back--reaction, averaging and the ``fitting problem'' \cite{BR1,BR2,BR3,BR4,CELU}, as well as the assumption that fulfilling the Copernican principle may require large scale homogeneity \cite{Copernican, chinos}. The Szekeres multi--structure configurations allow for looking at these issues under the assumption of a spacetime inhomogeneity that is far less idealised (and more related to the observed structures) than the very constrained spherical inhomogeneity of LT models that is normally employed in theoretical studies. 
\item {\bf The Zeldovich approximation.} The pancake morphology of over--densities (as those displayed in figures \ref{density} and \ref{pancake}) are reminiscent of structures that can be obtained in a Newtonian context through the Zeldovich approximation \cite{Zeld1}. In fact, the generalisation of the latter in General Relativity \cite{Zeld2,Zeld3,Zeld4} realises the axially symmetric particular case of exact Szekeres models of class II (see \cite{kasai,meures}). Since the Zeldovich approximation is an important conceptual tool in the study of structure formation (even in n--body numerical simulations \cite{Zeld5,Zeld6}), it is important to further explore these theoretical connections also for the models of class I that we have examined.      
\item {\bf Structure growth and red shift distortion.} By realising a coarse grained fitting of observed large scale structure (galactic clusters, superclusters and void regions) we can achieve a non--perturbative and fully relativistic approach to generalise previous work on these issues \cite{Linder1,Linder2,RSD}.
\item {\bf Peculiar velocities, Hubble flow and tidal forces.} Since the locations of the spatial maxima and minima of the density are not comoving (see figure 3 of \cite{nuevo} and figure \ref{LocHomSph}), we can define and study the peculiar velocity field of over--densities (along the lines of \cite{Zeld3}) with respect to the Hubble flow that follows from the level surfaces and spatial extrema of the Hubble scalar $\HH$ and the shear eigenvalue $\Sigma$, while the local field of tidal forces can be examined though the level surfaces and extrema of the eigenvalue $\Psi_2$ of the electric Weyl tensor.
\item {\bf Fitting observations without dark energy.} Large LT spherical void models used to verify observations fitting without assuming dark energy or cosmological constant have been practically ruled out because of the stringent restrictions placed by the Kinematic Sunyaev Zeldovich effect \cite{zibin1,zibin2,bull}. These restrictions can be re--examined by testing this effect on a non--spherical large scale geometry based on multiple structures generated by Szekeres models (as suggested in \cite{finns,chinos}). This would improve on previous work using Szkeres models for this purpose \cite{Bole2009-cmb,IRGW2008,BoCe2010,Buckley} that were axially symmetric or too idealised.
\item {\bf Limitations of the models.} While Szekeres dust configurations are ideal to describe non--trivial CDM sources in cosmological scales, they fail to  incorporate the effects of other forms of cosmic matter--energy sources, such as: photon radiation (which is not negligible in the early evolution stages near the last scattering time), or internal energy in warm dark matter models. Also, Szekeres configurations cannot incorporate the contribution of cosmic magnetic fields or gravitational waves, since the magnetic Weyl tensor is zero (they are Petrov type D solutions). However, all these effects can be examined by considering either Szekeres models admitting sources with non--zero pressure \cite{kras1,kras2}, or by adding the appropriate perturbative contributions associated to these sources to the dust source that defines the multiple structures we have constructed here.
\end{itemize}
We will undertake the study of these potential applications and extensions in future publications.

\section*{Acknowledgements.}

We acknowledge support from research grants DGAPA PAPIIT IA101414 and SEP-CONACYT 239639. IDG acknowldges support from CONACYT postgraduate grant program. 

\begin{appendix}

\section{Szekeres models in the standard representation.}\label{usual}

Szekeres models have been traditionally examined in the literature by means of the following metric (see its derivation and relation to various parametrisations in \cite{kras2}):
\ba \dd s^2 &=& -\dd t^2 +\frac{(\Phi'-\Phi\EE'/\EE)^2}{1 -K}\,\dd z^2+\frac{\Phi^2}{\EE^2}(\dd p^2+\dd q^2),\nonumber\\
\label{standard}\\
\EE &=&  \frac{S}{2}\,\left[1+\left(\frac{p-P}{S}\right)^2+\left(\frac{q-Q}{S}\right)^2\right],\label{Edef}\ea
where $\Phi=\Phi(t,z),\,\Phi'=\partial\Phi/\partial z$, the functions $K,\,P,\,Q$ and $S>0$ depend only on $z$ and the spatial coordinate ranges are $z\geq 0,\,\,-\infty < p,q <\infty$. The function $\Phi$ is determined by the Friedman--like equation   
\ba  \dot \Phi^2 &=& \frac{2M}{\Phi}-K+\frac{8\pi}{3}\Lambda \Phi^2,\nonumber\\
&\Rightarrow&\,\,\, t-\tbb=\int_0^\Phi{\frac{\sqrt{\bar\Phi} \dd\bar\Phi}{\left[2M - K\,\bar\Phi+ \frac{8\pi}{3}\Lambda \bar\Phi^3\right]^{1/2}}},\nonumber\\\label{cuadratura}
\ea
where $M=M(z)$ and $\tbb=\tbb(z)$ is the Big Bang time function. 

In order to relate (\ref{standard}) to the non--diagonal metric (\ref{g1})--(\ref{g3}), we choose the $z$ coordinate such that $\Phi_0=\Phi(t_0,z)=z$ holds for an arbitrary $t=t_0$. This choice allows for expressing $\tbb$ as a function of the free parameters $M,\,K,\,\Lambda$ from the quadrature in (\ref{cuadratura}) evaluated at $t_0$ (see \ref{tbb}). Then, we re--parametrise the metric functions as:
\begin{equation} a=\frac{\Phi}{z},\qquad \Gamma = \frac{z\Phi'}{\Phi},\qquad \bW =\frac{z\EE'}{\EE},\end{equation}
leading to the q--scalars
\ba
  \frac{4\pi}{3}\rho_q =\frac{M}{\Phi^3},\qquad \KK_q =\frac{K}{\Phi^2},\nonumber\\
   \HH_q =\frac{\dot\Phi}{\Phi}=\left[\frac{2M}{\Phi^3}-\frac{K}{\Phi^2}+\frac{8\pi}{3}\Lambda\right]^{1/2},\nonumber\ea
so that the fluctuations $\DDa$ and the scalars $A=\rho,\,\HH,\,\KK$ can be computed from (\ref{Adef}), (\ref{SigE}) and (\ref{AqDDa}). All initial value functions follow readily from the obtained expressions by setting $\Phi_0=z$ and $\Phi'_0=1$. The metric (\ref{g1})--(\ref{g3}) is finally obtained by the following stereographic coordinate transformation (see details in \cite{kras2})
\ba  p &=& P( r)+ S( r)\, {\rm cot} \left( \theta/2 \right) \cos (\phi),\nonumber\\
 q &=& Q( r) + S( r)\, {\rm cot} \left( \theta/2 \right) \sin (\phi),\nonumber\\
  z &=& r,\nonumber\\
 \label{stereo}\ea
which yields the form of $\bW$ in (\ref{dipole}) with the functions $X,\,Y,\,Z$ defined as
\begin{equation}X= \frac{zP'}{S},\qquad Y = \frac{zQ'}{S}, \qquad Z=\frac{zS'}{S}.\end{equation}
a re--parametrisation that is justified because the functions $S,\,P,\,Q$ only appear in the forms above in the metric (\ref{g1})--(\ref{g3}) and in the covariant scalars (\ref{Adef})--(\ref{SigE}).

All the results obtained in this article can be fully reproduced in the metric representation (\ref{standard}) by applying the transformations given above. In particular, the conditions (\ref{localhom1ab}) that define Local Homogeneity Spheres (see section \ref{PLH}) consist in demanding that $(4\pi/3)\rho'_{q0}=(M/z^3)'$ and $\KK'_{q0}=(K/z^2)'$ have common zeroes in a sequence of increasing coordinate values $z=0,z_*^1,..,z_*^n$.

\section{Regularity conditions.}\label{regularity}

\subsection{Regularity at the origin.}\label{origin}

The following regularity conditions must hold at $r=0$ for all $t$:
\ba  \left[A'_{q}\right]_c &=&\left[\Altb'\right]_c=\left[A'\right]_c=0,\,\,\Rightarrow\,\, 
     \left[\DDa\right]_c =\left[\DDaltb\right]_c=0,\nonumber\\ 
    &\Rightarrow&\,\,\left[A_{q}\right]_c =\left[\Altb\right]_c=\left[A\right]_c,\nonumber\\
     \left[\Sigma\right]_c &=& \left[\Sigltb\right]_c=\left[\Psi_2\right]_c=\left[\Pltb\right]_c=0,
    \label{origin1}\\
 \left[\,a\,\right]_c &>& 0,\quad \left[\,a'\,\right]_c=0, \quad \left[\,\Gamma\,\right]_c=1,\label{origin2}\\
 \left[X\right]_c &=& \left[X'\right]_c=\left[Y\right]_c=\left[Y'\right]_c=\left[Z\right]_c=\left[Z'\right]_c =0, \nonumber\\
  \W_c &=& \left[\bW\right]_c = 0,\label{origin3}\ea
where $[\,\,]_c$ denotes evaluation at $r=0$. In particular, we have for the dipole parameters $X,Y,Z\sim O(r^k)$ for $k>1$ in any neighborhood $r\approx 0$ \cite{kras2}. These regularity conditions are contained in (\ref{localhom}), which justifies describing the origin worldline as the Comoving Homogeneity Sphere of zero area (see section \ref{PLH}).

\subsection{Avoidance of shell crossing singularities.}\label{noshx}

Since we are only considering models that admit a single origin worldline (see Appendix D of \cite{sussbol}), the regularity conditions to prevent shell crossing singularities are \cite{kras2,sussbol}
\ba  \Gamma &>& 0\quad \hbox{(necessary)},\nonumber\\
  \Gamma - \bW &>& 0\quad \hbox{(necessary and sufficient)}.\nonumber\\
 \label{noshx1}\ea
Evidently, condition (\ref{noshx1}) restricts the dipole parameters $X,\,Y,\,Z$. Since the extrema (angular and radial) of $\bW$ lie in the curves $\B_\pm(r)$, then as a consequence of (\ref{WAE}) and (\ref{WAEmax})--(\ref{WAEmin}) the following conditions 
\ba \Gamma  &>& 0\quad\hbox{(necessary)},\nonumber\\
    \Gamma \mp\W(r) &>& 0\quad\hbox{(necessary and sufficient)}.\nonumber\\\label{noshx2}\ea     
are equivalent to (\ref{noshx1}) but easier to handle because they only depend on $(t,r)$. Also,  condition (\ref{noshx2}) evaluated at the initial time slice $t=t_0$ (where $\Gamma_0=1$) becomes:
\begin{equation} 1\mp \W=1\mp\sqrt{X^2+Y^2+Z^2}>0\label{noshx3}\end{equation}
which allows us to use absence of shell crossings to constrain the dipole parameters. 

Condition (\ref{noshx1}) (and its equivalent forms (\ref{noshx2}) and (\ref{noshx3})) also restrict the free functions of the LT seed model because of the presence of $\Gamma$. From its definition in (\ref{aGdef}), this metric function can be evaluated by implicit radial derivative of the quadrature of the Friedman equation (\ref{cuadratura}), which takes the following form in terms of q--scalars 
\ba\HH_{q0}(t-\tbb) &=& \nonumber\\
F(a,\Ommi,\OmLi)&\equiv& \int_0^a{\frac{\sqrt{\tilde a}\,\dd\tilde a}{\left[\Ommi-\Omki\,\tilde a+\OmLi\,\tilde a^3\right]^{1/2}}},\nonumber\\
\label{cuadratura2}\ea
where $\Omki=\Ommi+\OmLi-1$, with $\Ommi=8\pi\rho_{q0}/\HH_{q0}^2$ and $\OmLi=8\pi\Lambda/\HH_{q0}^2$, while the Big Bang time $\tbb$ is given in terms of these initial value functions by
\begin{equation}  \tbb= t_0-\frac{F_0}{\HH_{q0}}=\frac{1}{\HH_{q0}}\int_0^1{\frac{\sqrt{\tilde a}\,\dd\tilde a}{\left[\Ommi-\Omki\,\tilde a+\OmLi\,\tilde a^3\right]^{1/2}}}.\label{tbb}\end{equation}
where $F_0=F(1,\Ommi,\OmLi)$ follows from our choice of radial coordinate complying with $a_0=1$. 

For the case $\Lambda=0$ (and some cases with $\Lambda>0$) the quadrature (\ref{cuadratura2}), and thus $\Gamma$ and $\tbb$, can be obtained in terms of elementary functions, hence  (\ref{noshx1})--(\ref{noshx3}) can be simple analytic constraints between initial value functions (see \cite{sussbol} and in the standard variables in \cite{kras2}). For the general case $\Lambda>0$, (\ref{cuadratura2}) leads to hyper--elliptic integrals and thus (\ref{noshx1})--(\ref{noshx3}) needs to be evaluated numerically, with analytic constraints only possible in asymptotic evolution ranges $a\approx 0$ ($t\approx \tbb$) or $a\gg 1$ ($t\gg t_0$).

It is important to remark that (\ref{noshx1}) or even its simplified form (\ref{noshx2}) are far more stringent than the condition for absence of shell crossings in LT models: $\Gamma>0$. As a consequence, a Szekeres model may exhibit shell crossings even if its LT seed model is free from them.

\section{Rigorous classification of extrema of the Scalars.}\label{extrA}

At any fixed $t$ a scalar $A$ can be considered a real valued smooth function of $(r,\theta,\phi)$. The classification of its extrema as local maxima, minima or saddle points follows from the standard ``second derivative criterion'':
\begin{widetext}  
\ba  &{}& \hbox{\underline{Local minimum:}}
\qquad \hbox{det}\left(\left[\textrm{\bf{H}}_1\right]_{e}\right)>0,\quad \hbox{det}\left(\left[\textrm{\bf{H}}_2\right]_{e}\right)>0,\quad \hbox{det}\left(\left[\textrm{\bf{H}}\right]_{e}\right)>0,\label{mincond}\\
 &{}& \hbox{\underline{Local maximum:}}
\qquad \hbox{det}\left(\left[\textrm{\bf{H}}_1\right]_{e}\right)<0,\quad \hbox{det}\left(\left[\textrm{\bf{H}}_2\right]_{e}\right)>0,\quad \hbox{det}\left(\left[\textrm{\bf{H}}\right]_{e}\right)<0,\label{maxcond}\\
&{}&  \hbox{\underline{Saddle point:}\qquad\qquad any other combination of signs.}\label{saddle}
\ea
where ${}_e$ denotes evaluation at the extrema: points $(r_e,\theta_e,\phi_e)$ that satisfy (\ref{condAextr}), and $\textrm{\bf{H}},\,\textrm{\bf{H}}_2,\,\textrm{\bf{H}}_1$ are the Hessian matrix and its minors:  
\ba  \textrm{\bf{H}}=\left[ \begin {array}{ccc} A_{,{\theta\theta}}&A_{,{\theta\phi}}&A'_{,{\theta}}
\\ \noalign{\medskip}A_{,{\theta\phi}}&A_{,{\phi\phi}}&A'_{,{\phi}}
\\ \noalign{\medskip}A'_{,{\theta}}&A'_{,{\phi}}&A''\end {array}
 \right],\quad  \textrm{\bf{H}}_2=\left[ \begin {array}{cc} A_{,{\theta\theta}}&A_{,{\theta\phi}}\\ \noalign{\medskip}A_{,{\theta\phi}}
&A_{,{\phi\phi}}\end {array} \right],\quad \textrm{\bf{H}}_1=\left[A_{,{\theta\theta}}\right]. \label{Hessian}
\ea 
\end{widetext}
The second derivatives of $A$ evaluated at an extremum located in (\ref{condAextr}) are the angular derivatives (\ref{ang2derA1a}) together with the mixed radial--angular ones: 
\ba
 \left[A'_{,\phi}\right]_{e\pm} &=& \pm\frac{[\DDa]_e}{\Gamma_e\mp\W_e}\,\frac{Y'_eX_e-X'_eY_e}{\W_e},\nonumber\\
 \left[A'_{,\theta}\right]_{e\pm} &=& \frac{[\DDa]_e}{\Gamma_e\mp\W_e}\,\frac{(X_e^2+Y_e^2)Z'_e-\frac{1}{2}(X_e^2+Y_e^2)'Z_e}{\W_e\sqrt{X_e^2+Y_e^2}},\nonumber\\ 
\label{ang2derA1bc}\ea
and the second radial derivative  
\ba  \left[A''\right]_{e\pm} &=& \frac{(A_q)_e\Gamma_e}{\Gamma_e\mp\W_e}\left[(\alpha_\pm-\alpha_\pm^2)\da+\left(\da\right)''\right]_e,\label{radderA}\\
&\hbox{with:}&\quad \alpha_\pm \equiv \frac{3}{r}\left[(1+\da)\Gamma\mp\W\right]+\frac{\Gamma'}{\Gamma}-\frac{\Gamma'\mp\W'}{\Gamma\mp\W}.\nonumber\ea
where $\da=\DDaltb/A_q$ and (to simplify notation) the subindex ${}_e$ in the right hand sides denotes evaluation at either one of $r=r_{e\pm}$, while the symbol $\pm$ indicates evaluation at the curves $\B_\pm(r)$: for any quantity  $\CC$ the notation $\mp\CC_e$ is short hand for $-\CC(r_{e{+}})$ and $\CC(r_{e{-}})$. The determinant of the Hessian matrix and its two minors are
\ba \hbox{det}\left(\left[\textrm{\bf{H}}_1\right]_{e}\right) &=& \mp\,\frac{\W_e\,(\DDaltb)_e\,\Gamma_e}{(\Gamma_e\mp\W_e)^2},\label{detsHA1}\\
  \hbox{det}\left(\left[\textrm{\bf{H}}_2\right]_{e}\right) &=&  \frac{(X_e^2+Y_e^2)\Gamma_e^2\,(\DDaltb)^2_e}{(\Gamma_e\mp\W_e)^4}> 0,\label{detsHA2}\\ 
  \hbox{det}\left(\left[\textrm{\bf{H}}\right]_{e}\right) &=& \frac{(X_e^2+Y_e^2)\Gamma_e^2\,(\DDaltb)^2_e}{(\Gamma_e\mp\W_e)^2}\,\left[A''_e\pm (\DDaltb)_e\,\Psi_e\right],\nonumber\\\label{detsHA3}\ea
\begin{widetext}
\ba \hbox{with:}\quad \Psi_e \equiv \frac{\Gamma_e\left[(X'_eY_e-X_eY_e')^2\W_e^2+\left((X_e^2+Y_e^2)Z_e'-\frac{1}{2}(X_e^2+Y_e^2)'Z_e\right)^2\right]}{(\Gamma_e\mp\W_e)^2(X_e^2+Y_e^2)\W_e^3}>0,\nonumber\\
\label{Psi}\ea
\end{widetext} 
where $A''_e=\left[A''\right]_{e\pm}$. Since the determinant of $\textrm{\bf{H}}_2$ in (\ref{detsHA2}) is non--negative, we only need to evaluate the signs of the determinants of $\textrm{\bf{H}}_1$ and $\textrm{\bf{H}}$ in (\ref{detsHA1}) and (\ref{detsHA3}). Hence, the second derivative criterion (\ref{mincond})--(\ref{saddle}) can be stated at each extremum $\B_\pm(r_{e\pm})$ as
\ba  &{}&\hbox{local minimum:}\,\,h_\pm >0\,\,\hbox{and}\,\,\Delta_\pm>0,\label{mincond2}\\
     &{}&\hbox{local maximum:}\,\,h_\pm <0\,\,\hbox{and}\,\,\Delta_\pm<0,\label{maxcond2}\\
     &{}&\hbox{saddle: any sign combination such that}\,\,[h_\pm][\Delta_\pm]\leq 0,\nonumber\\\label{saddle2}\ea 
with $h_\pm$ and $\Delta_\pm$ given by
\ba h_\pm &\equiv& \hbox{sign}\left[\hbox{det}\left(\left[\textrm{\bf{H}}_1\right]_{e\pm}\right)\right]=\mp(\DDaltb)_e,\label{h1}\\
\Delta_\pm &\equiv& \hbox{sign}\left[\hbox{det}\left(\left[\textrm{\bf{H}}\right]_{e\pm}\right)\right]=A''_e\pm (\DDaltb)_e\,\Psi_e,\label{Delta}\ea
where we have assumed that $\Gamma_e>0$ and $\Gamma_e\mp\W_e>0$ hold from demanding absence of shell crossings.

Considering the signs of $A''_{e\pm}$, the nature of the extrema above is readily obtained from the following results: (i) the combination of signs of the determinants of $\textrm{\bf{H}}_1(A)$ and $\textrm{\bf{H}}(A)$ given by $h_\pm$ and $\Delta_\pm$ in (\ref{h1}) and (\ref{Delta}); and (ii) the concavity of the radial profiles $A_\pm$ yields the signs of the derivative $A''$ evaluated at the extrema $r_{e{+}},\,r_{e{-}}$ (see figure \ref{figprofiles}). Therefore, the following possibilities readily emerge from (\ref{mincond2})--(\ref{saddle2}) for $A$ depending on the concavity of the radial profiles $A_\pm$ in a given interval $\Delta_*^i$ (see  figure \ref{figprofiles} for reference):    
\begin{enumerate}
\item Extremum along $\B_{+}(r)$ with radial coordinate $r=r_{e{+}}^i>0$:
%\begin{enumerate}
%\item Local maximum at $r=0$ (central clump):
\begin{enumerate}
      \item If $(\DDaltb)_{e{+}}<0,\,A''_{e{+}}>0,\, h_+>0$ then $\Delta_+=|A''_{e{+}}|-|(\DDaltb)_{e{+}}|$. We have a local minimum if $\Delta_+>0$ and a saddle point if $\Delta_+\leq 0$.
      \item If $(\DDaltb)_{e{+}}>0,\,A''_{e{+}}<0,\, h_+<0$ then $\Delta_+=-|A''_{e{+}}|+|(\DDaltb)_{e{+}}|$. We have a local maximum if $\Delta_+<0$ and a saddle point if $\Delta_+\geq 0$.
\end{enumerate}
\item Extremum along $\B_{-}(r)$ with radial coordinate $r=r_{e{-}}^i>0$:
\begin{enumerate}
     \item $(\DDaltb)_{e{-}}<0,\, A''_{e{-}}>0,\, h_{-}<0$, hence $\Delta_{-}=|A''_{e{-}}|+|(\DDaltb)_{e{-}}|>0$, therefore only saddle points are possible.
     \item $(\DDaltb)_{e{-}}>0,\, A''_{e{-}}<0,\, h_{-}>0$, hence $\Delta_{-}=-|A''_{e{-}}|-|(\DDaltb)_{e{-}}|<0$, therefore only saddle points are possible. 
\end{enumerate}
\end{enumerate}
These results hold for extrema in any interval $\Delta_*^i$ for all combinations of patterns displayed by figure \ref{figprofiles}. This classification is easily extended to the extrema of $\Sigma,\,\Psi_2$ and to extrema that arise from simulated shell crossings without assuming PLH (see section \ref{noPLH}). The only extrema that are not covered by the criterion (\ref{mincond})--(\ref{saddle}) are those occurring at the $r_*^i$ where $\DDa=0$ holds (see panels (a) and (b) of figure \ref{figprofiles}). The fact that these extrema are saddles follows from looking at the behaviour of $A$ in a small neighbourhood around the points $(r_*^i,\theta_\pm(r_*^i),\phi_\pm(r_*^i))$. 

The rigorous classification of the extrema described above coincides with the qualitative one in section \ref{classi}, as expected from the specific sign combinations of $h_\pm$ and $\DDaltb$ at $\B_\pm(r)$. However, it follows from the form of $\Delta_\pm$ in (\ref{Delta}) that even along $\B_{+}(r)$ same radial and angular concavity and a radial maximum or minimum (sign of $A''_{e\pm}$) are only necessary (but not sufficient) conditions for a spatial maximum or minimum of $A$.

\section{Proof of the existence condition of extrema of the radial profiles.}\label{formalproofs}

Proposition 1 generalises the following known result for LT models (Lemma 3 of \cite{RadProfs} for a single interval $\Delta_*^1 = 0<r<r_*^1$): a zero of $A'_q$ at $r=r_*^1$ is sufficient for the existence of an extremum of the radial profile of its associated covariant LT scalar $\Altb$ in  $\Delta_*^1$. The proof follows from the integral form of (\ref{DDaltb}) obtained from the  integrability of the LT scalar $\Altb R^3$ with $R=a\,r$:
\begin{equation}\DDaltb = \Altb-A_q = \frac{1}{R^3}\int_0^r{\Altb'\,R^3\,\dd\bar r}\label{intDDazero}.\end{equation}
Considering the first interval $\Delta_*^1=0<r<r_*^1$ (see figure \ref{figprofiles}), we have from (\ref{DDaltb}) at any fixed $t$  
\ba 0 &=& \Altb(t,r_*^1)-A_q(t,r_*^1)\nonumber\\
 &=& \frac{1}{R^3(t,r_*^1)}\int_0^{r_*^1}{\Altb'(t,\bar r)\,R^3(t,\bar r)\,\dd\bar r},\label{intlema1}\ea
which implies that the integrand must change sign at least once in the interval $\Delta_*^1$. Since $R>0$ for $r>0$, then $\Altb'$ must have (at least) a zero at some value $r=\rtv$ in this interval. For $n$ intervals $\Delta_*^i$ associated with PLH we obtain this result by integrating along each interval separately.   

Proposition 1 is proven by generalising Lema 3 of \cite{RadProfs} to Szekeres scalars and for $n$ intervals $\Delta_*^i$. For this purpose the LT radial integral (\ref{intDDazero}) can be generalised for the Szekeres geometry in spherical coordinates as the following 3--dimensional integral
\ba  \int\limits_{\phi}\!\int\limits_{\theta}{
    {\dd\phi\,\dd\theta\,\left(A\,\Y^3\right)}
             } &=&  \int\limits_{\phi}\!\int\limits_{\theta}{
    {\dd\phi\,\dd\theta\,\int\limits_{r}{\dd r\,\left[A\,\Y^3\right]'}}}\nonumber\\
    &=& \int\limits_{\phi}\!\int\limits_{\theta}{
    {\dd\phi\,\dd\theta\,\int\limits_{r}{\dd r\,\left[A'\,\Y^3+A(\Y^3)'\right]}}},\nonumber\\
     \label{SzInt1}  \ea
where the angular variables $(\theta,\phi)$ take their usual full ranges and $\Y$ satisfies $r\Y'/\Y=\Gamma-\bW$. Considering that $A'_q=(3\Y'/\Y)(A-A_q)$ follows from (\ref{AqDDa}), we have $A(\Y^3)'=3\Y^2\Y'A_q+A'_q\Y^3=(A_q \Y^3)'$. Substitution into (\ref{SzInt1}) yields 
\ba  \int\limits_{\phi}\!\int\limits_{\theta}{
    {\dd\phi\,\dd\theta\,\left[(A-A_q)\,\Y^3\right]}
             } &=&  \DDaltb\Gamma\int\limits_{\phi}\!\int\limits_{\theta}{
    {\dd\phi\,\dd\theta\,\frac{\Y^3}{\Gamma-\bW}}
             }\nonumber\\
             &=& \int\limits_{\phi}\!\int\limits_{\theta}{
    {\dd\phi\,\dd\theta\,\int\limits_{r}{\dd r\,\left(A'\,\Y^3\right)}}}.\nonumber\\
     \label{SzInt2}  \ea
where in the second integral we used $rA'_q/3=(\Altb-A_q)\Gamma=\DDaltb\Gamma$. If we assume PLH and consider the first interval $\Delta_*^1$, then the left hand side of (\ref{SzInt1}) vanishes identically at $r=r_*^1$, hence
\begin{equation} 0 = \int\limits_{\phi}\!\int\limits_{\theta}{
    {\dd\phi\,\dd\theta\,\int\limits_{0}^{r_*^1}{\dd r\,\left(A'\,\Y^3\right)}}}\end{equation}  
which implies that $A'$ must have (at least) a zero in this interval at some $r=\rtv$ that smoothly depends  on the angular coordinates. In particular, there most be a zero of $A'$ at $\theta_\pm,\phi_\pm$.  The proof is easily extended to the $n$ intervals by integrating along them. 

\subsection{The converse of Proposition 1.}\label{converse}

Notice that the converse of Lema 3 of \cite{RadProfs} is false: an extremum of $\Altb$ may exist in a given radial range $r>0$ even without extrema of $A_q$ ({\it i.e.} even if we do not assume PLH initial conditions (\ref{localhom1ab}) so that $A'_q\ne 0$ holds everywhere in this range). The proof is as follows.  We assume that $A'_q<0$ holds in any given radial range $r>0$, together with the existence of a zero of $\Altb'$ at some $r=\rtv>0$, with $\Altb'(t,r)<0$ for $0<r<\rtv$ and $\Altb'(t,r)>0$ for $r>\rtv$ (the proof is analogous for the opposite signs). The integral (\ref{intDDazero}) for any upper integration limit $r>\rtv$ can be written as
\begin{equation} \DDaltb\,R^3= -\int_0^{\rtv}{\left|\Altb'\right|R^3 \dd \bar r}+\int_{\rtv}^r{\Altb'R^3 \dd \bar r}<0.\label{LTint2}\end{equation}
This inequality is not {\it a priori} contradictory: it can hold if the profile of $\Altb$ sharply decays for $r<\rtv$ (for $r\approx\rtv$) and remains nearly constant for $r>\rtv$, hence the first integral above can be sufficiently large and negative and the second one (which is positive) sufficiently small to keep the negative sign of $A'_q$. This result can be easily generalised to Szekeres scalars. 

\end{appendix}

\end{document}